\documentclass[10pt,a4paper]{article} 
\usepackage{graphicx}
\usepackage{style}
\usepackage{epstopdf}
\usepackage{epsfig}
\usepackage{amsmath,amssymb,bm,amsfonts,yfonts,bbm}
\usepackage{appendix}

\newcommand{\be}{\begin{equation}}
\newcommand{\ee}{\end{equation}}
\newcommand{\bea}{\begin{eqnarray}}

\newcommand{\eea}{\end{eqnarray}}

\newcommand{\ex}{\epsilon}

\newcommand{\kx}{\kappa}

\def\Rc{{\cal R}}
\def\Sc{{\cal S}}
\def\Fc{{\cal F}}

\newcommand{\Rb}{\bar{R}}

\newcommand{\kt}{{\tilde{k}}}

\newcommand{\etapz}{\eta_{\perp 0}}

\def \lta {\mathrel{\vcenter
     {\hbox{$<$}\nointerlineskip\hbox{$\sim$}}}}
\def \gta {\mathrel{\vcenter
     {\hbox{$>$}\nointerlineskip\hbox{$\sim$}}}}

\makeatletter
\renewcommand{\citation}[1]{%
  \g@addto@macro{\citation@list}{,#1}%
}
\newcommand*{\citation@list}{} 
\newcommand{\sortbibitem}[2]{%
  \global\@namedef{bibitem@#1}{%
    \bibitem{#1} #2
  }%
}
\newcommand{\sort@bibitems}{%
  \@for\next:=\citation@list\do{%
    \@nameuse{bibitem@\next}%
    \global\@namedef{bibitem@\next}{}%
  }%
}
\expandafter\def\expandafter\endthebibliography\expandafter{%
  \expandafter\sort@bibitems\endthebibliography
}
\makeatother

\DeclareSymbolFont{mathscrUC}{U}{rsfs}{m}{n}  
\DeclareSymbolFont{mathscrLC}{OT1}{pzc}{m}{n} 

\makeatletter

\DeclareRobustCommand*{\mathscr}[1]{\gdef\F@ntPrefix{mathscr@char@}%
  \@EachCharacter #1\@EndEachCharacter}
\long\def\DoLongFutureLet #1#2#3#4{%
   \def\@FutureLetDecide{#1#2\@FutureLetToken
      \def\@FutureLetNext{#3}\else
      \def\@FutureLetNext{#4}\fi\@FutureLetNext}
   \futurelet\@FutureLetToken\@FutureLetDecide}
\def\DoFutureLet #1#2#3#4{\DoLongFutureLet{#1}{#2}{#3}{#4}}
\def\@EachCharacter{\DoFutureLet{\ifx}{\@EndEachCharacter}%
   {\@EachCharacterDone}{\@PickUpTheCharacter}}
\def\m@keCharacter#1{\csname\F@ntPrefix#1\endcsname}
\def\@PickUpTheCharacter#1{\m@keCharacter{#1}\@EachCharacter}
\def\@EachCharacterDone \@EndEachCharacter{}

\makeatother

\relax
\title{The effect of multiple features on the power spectrum in 
two-field inflation}

\author{K. Boutivas, I. Dalianis, G.P. Kodaxis and N. Tetradis}

\affiliation{Department of Physics, University of Athens, University Campus, Zographou 157 84, Greece}

\emailAdd{kboutivas@phys.uoa.gr, idalianis@phys.uoa.gr, gekontax@phys.uoa.gr, ntetrad@phys.uoa.gr}

\abstract{
We extend our previous work 
on the enhancement of the curvature spectrum during inflation to the two-field case.
We identify the slow-roll parameter $\eta$ as the quantity
that can trigger the rapid growth of perturbations.  
Its two components, $\eta_\parallel$ along the background trajectory
and $\eta_\perp$ perpendicular to it, remain small during most of the evolution, 
apart from  short intervals during which they take large, positive or negative, values.
The typical reason for the appearance of strong features in $\eta_\parallel$ is
sharp steps or inflection points in the inflaton potential,
while $\eta_\perp$ grows large during sharp turns in
field space. We focus on the additive effect of several features
leading to the resonant growth of the curvature spectrum. 
Three or four features in the evolution of
 $\eta$ are sufficient in order to induce an enhancement of the 
 power spectrum by six or seven orders of magnitude, which can
lead to the significant production of primordial black holes and stochastic 
gravitational waves.
A big part of our study focuses on understanding the evolution
of the perturbations and the resulting spectra through analytic means.
The presence of multiple features in
the background evolution points to a more complex inflationary paradigm, 
which is also more natural in the multi-field case. 
The critical examination of this possibility is within 
the reach of experiment.
}

\begin{document}

\maketitle

\section{Introduction}

In the theory of cosmic inflation the cosmological perturbations  
originate from the quantum fluctuations of the fields that drive inflation \cite{Mukhanov:1981xt, Hawking:1982cz, Starobinsky:1982ee, Guth:1982ec}.  
The quantum fluctuations of the Bunch-Davies vacuum 
are stretched at macroscopic scales beyond the Hubble radius and become the 
seeds for the growth of structure in the universe after inflation ends. 
The evolution of the fluctuations on the quasi-de Sitter background is 
affected by the deviations from scale invariance. Strong features in the 
background can leave characteristic imprints on the spectrum of perturbations
with observable consequences. 
The CMB anisotropies and the large-scale structure of the universe offer us a window to  the profile of the spectrum of the primordial perturbations 
generated in the early stages of inflation \cite{Planck:2018jri}. 
Perturbations generated during the later stages 
are not imprinted on the CMB sky. However, such small-scale 
inhomogeneities  might be of particular importance if their amplitude is large. 
A web of strong density perturbations generates a stochastic gravitational 
wave (GW) background, potentially observable by operating or designed experiments
that cover different frequency bands,  ranging from  nanohertz \cite{skate, ipta, Hobbs:2009yy, Janssen:2014dka, Lentati:2015qwp, NANOGrav:2020bcs} to
milihertz and decihertz  \cite{elisa, LISA:2017pwj, Barausse:2020rsu,  Kawamura:2020pcg, Ruan:2018tsw, TianQin:2015yph, Badurina:2019hst}, and up to $10^3$ hertz \cite{KAGRA:2021kbb, Maggiore:2019uih}.  
Overdensities with amplitudes beyond a particular threshold collapse into black holes 
that contribute to the dark matter density.  
Interestingly enough, if these black holes are in the stellar-mass region, 
they may be probed by the LIGO-Virgo-KAGRA GW experiments.

The nature of the inflaton field that generates the 
perturbations remains elusive. Actually, it might be the case that the 
inflationary phase is implemented by more than one fields. Both single 
and multi-field inflation models can describe well the cosmological 
data of primordial perturbations, assuming that  the latter  make their 
multi-field dynamics manifest at scales smaller than those directly observed 
in the CMB sky.
The phenomenology of interest in multi-field inflation models is related to the 
fact that the curvature 
perturbations can evolve even on super-Hubble scales because of the presence 
of isocurvature perturbations  
\cite{Starobinsky:1994mh, sasaki, Garcia-Bellido:1995hsq, Linde:1996gt, 
Langlois:1999dw, Gordon:2000hv, Tsujikawa:2000wc},  
called also entropy perturbations or 
non-adiabatic pressure perturbations \cite{Linde:1985yf, Kofman:1985zx, 
 Polarski:1994rz,  Mukhanov:1997fw}.
In particular inflationary set-ups, the evolution of the curvature perturbations 
triggered by isocurvature modes can be dramatic, generating  an observable GW signal and potentially a significant primordial black hole (PBH) abundance \cite{Pi:2017gih, Palma:2020ejf, Fumagalli:2020nvq, Braglia:2020taf}.
At the same time, the isocurvature modes can be absent at the CMB scales 
$\gtrsim 1$ Mpc, where the curvature perturbations are effectively described by single-field results.

From a particle-physics point of view, it is natural to expect  
multi-field dynamics during inflation.  When more than one fields are rolling,  
one can define an adiabatic perturbation component along the  direction tangent to 
the background classical trajectory, and isocurvature perturbation components 
along the directions orthogonal to the trajectory\cite{Gordon:2000hv}. 
Curvature perturbations may be affected by the isocurvature
perturbations if the background solution follows a curved trajectory 
in field space. 
Models of inflation with curved inflaton trajectories have been studied often 
in the past \cite{Achucarro:2010da, Achucarro:2010jv, Chen:2011zf, Shiu:2011qw, Cespedes:2012hu, Avgoustidis:2012yc, Gao:2012uq, Achucarro:2012fd, Konieczka:2014zja, Brown:2017osf, Achucarro:2019pux, Fumagalli:2019noh, Bjorkmo:2019fls, Christodoulidis:2019jsx, Braglia:2020eai, Aldabergenov:2020bpt, Palma:2020ejf, Fumagalli:2020adf, Fumagalli:2020nvq, Iacconi:2021ltm}.
When the inflaton slow-rolls along a deep valley while the mass term perpendicular 
to the trajectory is large, the isocurvature perturbation can be integrated out.
This introduces corrections to the effective single-field theory, 
which can be absorbed in the sound speed for the low-energy 
perturbations \cite{Tolley:2009fg, Cremonini:2010ua,  Achucarro:2012sm, Pi:2012gf, Achucarro:2012yr}.
In the particular case of a sharp turn in field space  
\cite{Achucarro:2010da, Achucarro:2010jv, Chen:2011zf, Shiu:2011qw, Cespedes:2012hu, Avgoustidis:2012yc, Gao:2012uq, Achucarro:2012fd, Konieczka:2014zja},  
the impact of the isocurvature modes is more prominent, with the resulting
curvature spectrum departing significantly from scale invariance. 
An oscillatory pattern emerges  
at  wavenumbers  characteristic of the turn \cite{Shiu:2011qw, Cespedes:2012hu}.  
Also, a significant amplification takes place if during the sharp turn 
the isocurvature modes experience a transition from heavy to light 
\cite{Palma:2020ejf, Fumagalli:2020adf, Fumagalli:2020nvq,   Fumagalli:2021mpc}. 
For broader turns the amplification is still possible \cite{Braglia:2020eai, Aldabergenov:2020bpt}, but the oscillatory patterns fade away.

In this paper we extend our previous work by going beyond the single-field inflationary 
scenario.  
We concentrate on the case of two fields that experience strong features in the course of 
the inflationary evolution. With the term feature we mean sharp deviations 
from the minimal and smooth inflationary evolution that have 
observable consequences, such as distinctive signatures in the 
power spectra of perturbations, see for example 
refs. \cite{Chen:2010xka, Chluba:2015bqa, Slosar:2019gvt}. 
 The origin of the features is attributed to the intrinsic and complex dynamics of the 
 system of inflaton fields. 
 Aiming at a general and model-independent analysis,  
we follow an effective description, parameterizing the  dynamics 
behind the strong features via the background evolution encoded in  the 
slow-roll parameters. This phenomenological  description  
also helps to maintain a geometrical  intuition about the field space trajectory.
Suggestive constructions that capture the dynamics of the subtle underlying mechanisms 
that produce these features can be inversely engineered. 
Particular set ups such as those of  refs. \cite{DAmico:2020euu, Spanos:2021hpk, Iacconi:2021ltm}, just to mention a few,  indicate some model building directions.

We focus on two particular kinds of strong dynamical features, arising either from 
sharp steps in the potential or 
turns in the inflationary trajectory. 
Such brief but strong departures from a steady slow-roll evolution act as a 
source for the curvature perturbations, which can be amplified significantly. 
Even though the two features seem quite distinct,
they have a similar description at the level of the slow-roll parameters. 
This becomes apparent if we decompose, along with the perturbations,  
the second slow-roll parameter into its tangent $\eta_{\parallel}$ and 
orthogonal $\eta_{\perp}$ components. Sharp steps lead to large positive
values of $\eta_\parallel$, while sharp turns result in large values of 
$\eta_\perp$, with either sign. 
We examine each case separately. Allowing both $\eta_\parallel$ and $\eta_\perp$
to become large leads to a combination of effects, but no new qualitative behaviour.   
When the leading effect comes from the rapid change of 
the $\eta_{\parallel}$ component, the inflationary dynamics 
is effectively  described by a single-field theory. A step-like transition 
in the inflaton potential energy can cause this change. We studied
this possibility in previous work, following mainly a numerical approach  \cite{Kefala:2020xsx,Dalianis:2021iig}. 
We extend this work here by focusing on the analytic understanding 
of the evolution of the perturbations and the resulting spectra.
In the second case we study, 
the $\eta_{\perp}$ component is rapidly changing when a turn in field 
space takes place.  In this case it is the isocurvature perturbations  
that get excited. Through their coupling to the curvature perturbations, which is
proportional to $\eta_\perp$, they play a crucial role in shaping the final 
curvature spectrum.

In the case of large $\eta_\parallel$, we calculate the power spectrum of 
curvature perturbations following a semi-analytic approach.  
The conventional perturbative approach has limitations because 
the equations of motion can be solved 
analytically only when the slow-roll parameters have 
a mild time dependence within the slow-roll approximation.
We utilize here the Green's function method in order to recast the system 
of equations of motion for the  fields in a form that admits an iterative solution. 
We also find a formula  that gives an approximate solution in closed form.  
For a moderate enhancement of the slow-roll parameter our 
formalism can give a good quantitative description of the evolution of 
the curvature perturbation. 
This formalism can also describe other features, such as an inflection point in the
inflaton potential. It can be extended to the two-field case, even though the 
resulting expressions are rather complicated. 
The semi-analytic approach has several advantages over a purely numerical treatment:
one can read off characteristic frequencies, analyze non-minimal initial conditions in a 
straightforward manner, and reduce computational time. 
However,  the quantitative accuracy is limited to  enhancements of 
the curvature power spectrum up to three or four orders of magnitude.  
For larger enhancements our approximation captures well the characteristic frequencies,
but not the exact size of the amplification. 
In this case the numerical solution for the system of equations of motion is necessary. 
 
For the two-field inflation system with a transient large value of $\eta_\perp$,
a strong enhancement of the curvature spectrum is possible. 
Previous studies have concluded that an enhancement by several orders of magnitude
can be achieved only for a nonzero curvature of the internal manifold spanned by the
fields \cite{Palma:2020ejf,Fumagalli:2020nvq}. This would require 
noncanonical kinetic terms. We show that a large enhancement can
take place even for vanishing internal
curvature if there are several instances during which  $\eta_\perp$ grows large.
Such a situation would occur if, for example, 
the two-field system experiences several sharp
turns during its evolution. The mechanism for the enhancement is different from
the case of a single large-$\eta_\perp$ event and leads
to spectra with a different profile. 
For this reason our analysis and results differ from previous studies 
in this subject,
in which the assumption of a non-trivial field-space metric was the essential 
ingredient.
More specifically, we find that a 
system of two fields with canonical kinetic terms can source large curvature 
perturbations if there are a few turns in the background trajectory, 
each taking place in a time interval less than a Hubble time. 
We consider turns smaller than $\pi$ with alternating signs, so that the 
field trajectory does not cross itself. Three or four turns  are sufficient 
to enhance the curvature power spectrum by a factor of $10^7$.  
A larger number of rather small but sharp turns can produce an equally large enhancement.
We assume that these features  occur at intermediate or late stages 
of the inflationary evolution, so that the spectrum in the CMB range remains unaffected.

The fact that
a prominent peak in the density spectrum is generated implies that a 
strong GW signal is also induced at the moment of horizon reentry. 
The GW channel is a portal to the 
primordial density perturbations at small scales and can probe the inflationary dynamics.
Hence, along with the excited curvature power spectra we  present the 
corresponding tensor power spectra induced at second order in perturbation theory
\cite{Mollerach:2003nq, Ananda:2006af, Baumann:2007zm, Saito:2008jc}. 
 We find that at least three turns are required in order for the induced GW signal to be 
 strong enough to be detected.
Large fluctuations of the curvature perturbation can trigger gravitational 
collapse, which, though an extremely rare event, becomes important 
above a particular threshold \cite{Carr:2020gox}.
For completeness we consider the corresponding PBH production for two  benchmark mass values and PBH  abundances that can be tested by current and near future GW experiments.

The paper is organized as follows: In section \ref{twofieldsection} we present the background evolution of the two-field inflaton system, as well as the evolution of
the cosmological perturbations, including the interaction between 
isocurvature and curvature modes. 
 In section \ref{notes} 
 we present approximate analytic solutions for the first class of strong features that
 result from steps and inflection points in the potential.
 In section \ref{sectionturns} we analyze the second class of features that 
 result from sharp turns in field space.  
 In section \ref{GWPBH} we present the tensor spectra and the PBH counterpart, 
 focusing on the case in which the final curvature perturbation originates almost exclusively in the interactions with the isocurvature modes.
In section \ref{conclusions} we present our conclusions. 
In appendix \ref{appendixa} we provide an assessment of the accuracy
of analytical estimates of the spectrum  presented in the main text.
In appendix \ref{appendixb} we give some detailed expressions
resulting from the application of
the Green's function method to the two-field system.

\noindent


\section{Cosmological perturbations in two-field inflation} \label{twofieldsection}

\subsection{Background evolution}

The action is of the form
\be
S=\int d^4x\,\sqrt{-g} \left[\frac{1}{2}R
-\frac{1}{2} g^{\mu\nu}\gamma_{ab} \partial_\mu\phi^a\partial_\nu\phi^b
-V(\phi) \right],
\label{actiontwo} \ee
with $a=1,2$.
We have chosen units such that the Planch mass is $M_{\rm Pl}=1$.
On an expanding, spatially flat background, with scale factor $a(t)$,
the equations of motion of the background fields take the form \cite{Achucarro:2012yr, Cespedes:2012hu}
\begin{eqnarray}
\frac{D}{dt}\dot{\phi}^a+3H\dot{\phi}^a+V^a&=&0
\label{back1} \\
3H^2&=&\frac{1}{2}\dot{\phi}^2+V,
\label{back2} \end{eqnarray}
where $V^a=\gamma^{ab}\partial V/\partial \phi^b$, $H=\dot{a}/{a}$, with a dot denoting
a derivative with respect to cosmological time, and
\be
\frac{D}{dt}X^a=\dot{X}^a+\Gamma^{a}_{bc}\,\dot{\phi^b}\,X^c.
\label{covD} \ee
By defining $\dot{\phi}^2\equiv \gamma_{ab}\dot{\phi}^a\dot{\phi}^b$, 
we obtain
\be
\dot{H}=-\frac{\dot{\phi}^2}{2}.
\label{raych} \ee

We next define vectors $T^a$ and $N^a$ tangent and normal to the path
\begin{eqnarray}
T^a&=&\frac{\dot{\phi}^a}{\dot{\phi}}
\label{TT} \\
N_a&=&(\det \gamma)^{1/2}\epsilon_{ab}T^b,
\label{NN}
\end{eqnarray}
such that $T^aT_a=N^aN_a=1$, $T^aN_a=0$. 
Projecting eq. (\ref{back1}) along $T^a$, one finds
\be
\ddot{\phi}+3H\dot{\phi}+V_T=0,
\label{friedmann} \ee
where $V_T=T^a\partial V/\partial \phi^a$. 
One also finds
\be
\frac{DT^a}{dt}=-\frac{V_N}{\dot{\phi}}N^a,
\label{dTdt} \ee
with $V_N=N^a\partial V/\partial \phi^a$.
The slow-roll parameters are defined as 
\begin{eqnarray}
\epsilon\equiv -\frac{\dot{H}}{H^2}=\frac{\dot{\phi}^2}{2H^2}
\label{slowep} \\
\eta^a\equiv -\frac{1}{H\dot{\phi}}\frac{D\dot{\phi}^a}{dt}.
\label{slowet} 
\end{eqnarray}
Then $\eta ^a$ can be decomposed as
\be
\eta^a=\eta_\parallel T^a+\eta_\perp N^a,
\label{decomp} \ee
with 
\begin{eqnarray}
\eta_\parallel&=&-\frac{\ddot\phi}{H\dot{\phi}}=-\frac{\dot{\epsilon}}{2H\epsilon}
+\epsilon
\label{etaa} \\
\eta_\perp&=&\frac{V_N}{H \dot{\phi}}.
\label{etab} \end{eqnarray}
We also have
\begin{eqnarray}
\frac{DT^a}{dt}&=&-H\eta_\perp N^a
\label{Neta} \\
\frac{DN^a}{dt}&=&+H\eta_\perp T^a.
\label{Teta} \end{eqnarray}

\subsection{Perturbations}

The evolution equations for the curvature and isocurvature perturbations 
in two-field inflation can be cast in the form \cite{Achucarro:2012yr, Cespedes:2012hu}
\begin{eqnarray}
\Rc_{k,NN}+\left(3+\epsilon-2\eta_\parallel \right)\Rc_{k,N}+
\frac{k^2}{H^2}e^{-2N}\Rc_k&=&-2\frac{\eta_\perp}{\sqrt{2\epsilon}}\left[ 
\Fc_{k,N}+(3-\eta_\parallel-\xi_\perp)\Fc_k \right]
\nonumber \\
&~&~
\label{RR} \\
\Fc_{k,NN}+ (3-\epsilon)\Fc_{k,N}+
\frac{k^2}{H^2}e^{-2N}\Fc_k+\left(\frac{M^2}{H^2}+\epsilon\,{\mathbb R}
-\eta_\perp^2 \right)\Fc_k&=&
2\sqrt{2\epsilon}\, \eta_\perp \Rc_{k,N},
\label{FF} \end{eqnarray}
with  
\begin{eqnarray}
\eta_\parallel&=&\epsilon-\frac{\epsilon_{,N}}{2\epsilon}
\label{epsN} \\
\xi_{\perp}&=&-\frac{{\eta}_{\perp,N}}{\eta_\perp}.
\label{xiN} \end{eqnarray}
We have written the equations in Fourier space, using the number of efoldings
$N$ as independent variable. 
The subscripts denote derivatives with respect to $N$.
Here $\Rc_k$ is the curvature perturbation, while
$\Fc_k$ is related to the isocurvature perturbation $\Sc$ through 
$\Fc_k=\sqrt{2\epsilon}\,\Sc_k$. The mass $M$ of the isocurvature perturbation is
given by the curvature of the potential in the direction perpendicular to the
trajectory of the background inflaton. The variable ${\mathbb R}$
is the Ricci scalar of the  
internal manifold spanned by the scalar fields \cite{sasaki}.
It vanishes for a model with standard kinetic terms for the two fields. 

In order to focus on the main features associated with the enhancement of
the curvature spectrum, we make some simplifying assumptions:
\begin{itemize}
\item 
We approximate the Hubble parameter as constant. This is a good approximation,
as its variation during the period of interest is ${\cal O}(10\%)$, while 
the spectrum may increase by several orders of magnitude. 
\item
In a similar vein, we 
take the mass $M$ of the isocurvature modes to be constant. 
We also assume that $M\gta H$, so that the
isocurvature perturbations are suppressed apart from short periods during which the 
parameter $\eta_\perp$ becomes large. 
\item 
We do not
consider the possibility of a curved field manifold, but assume that the 
fields have standard kinetic terms. This means that we can set ${\mathbb R}=0$.
\item
We assume that the parameter $\epsilon$ takes a small constant value while
the system is in the slow-roll regime, consistently with the constraints from
the cosmic microwave backgroung (CMB). We neglect here corrections arising 
from the slow-roll regime that
lead to small deviations from scale invariance. We focus instead on
short periods of the inflaton evolution during which $\eta_\parallel$ or $\eta_\perp$
can grow large.
These periods are reflected in strong  
deviations of the spectrum from scale invariance over a range of momentum scales.

\end{itemize}

We are interested in strong deviations from the slow-roll regime during short intervals
in $N$, which can result in the strong enhancement of the curvature perturbations. 
There are two typical scenaria that we have in mind:
\begin{itemize}
\item
For $\eta_\perp^2 \ll M^2/H^2$, the isocurvature mode is strongly suppressed and
the rhs of eq. (\ref{RR}) vanishes. The curvature mode can be enhanced if the 
coefficient of the term $\sim \Rc_{k,N}$ in the lhs becomes negative. This 
requires large positive values of the parameter $\eta_\parallel$. Such values can
be attained if the inflaton potential
displays an inflection point, or a sharp step \cite{Kefala:2020xsx, Dalianis:2021iig}.
The dominant effect comes from $\eta_\parallel$,
which can take very large values, while $\epsilon$ takes values at most around 1
and can be neglected.
The equation for the curvature perturbation can be approximated as 
\be
\Rc_{k,NN}+( 3-2\,\eta_\parallel)\, \Rc_{k,N}+\frac{k^2}{H^2}e^{-2N}\Rc_k=0.
\label{RfN} \ee
\item
If $\eta_\perp^2 \gg M^2/H^2$ for a short period, the isocurvature modes 
can be temporarily excited very strongly. The rhs of eq. (\ref{RR}) then becomes
large and acts as a source for the curvature perturbations, leading to their strong
enhancement. At a later time, $\eta_\perp$ becomes small and the isocurvature
perturbations become suppressed again. This process can take place while
the slow-roll parameters $\epsilon$ and $\eta_\parallel$ remain small.
In order to capture the essence of this mechanism, we assume that $\epsilon$ is 
small and roughly constant, and switch to the field $\Sc_k=\Fc_k/\sqrt{2\epsilon}$.
The system of eqs. (\ref{RR}), (\ref{FF}) becomes
\begin{eqnarray}
\Rc_{k,NN}+3\Rc_{k,N}+
\frac{k^2}{H^2}e^{-2N}\Rc_k&=&-2\left( 
\eta_\perp\Sc_{k,N}+
\eta_{\perp,N}\Sc_k+3\eta_\perp \Sc_k \right)
\nonumber \\
&~&~
\label{RRR} \\
\Sc_{k,NN}+ 3\Sc_{k,N}+
\frac{k^2}{H^2}e^{-2N}\Sc_k
+\left(\frac{M^2}{H^2} -\eta_\perp^2 \right)\Sc_k&=&
2\, \eta_\perp \Rc_{k,N}.
\label{SSS} \end{eqnarray}
Notice that $\Sc_k$ is the appropriate field on which the initial condition of
a Bunch-Davies vacuum can be imposed, similarly to $\Rc_k$.
\end{itemize}
In both the above scenaria, it is the acceleration in the evolution of the
background inflaton, either in the direction of the flow through $\eta_\parallel(N)$, or perpendicularly to it through $\eta_\perp(N)$,
that causes the amplification of the spectrum of 
curvature perturbations. Another common characteristic is that 
the spectrum displays strong oscillatory patterns.

A weak point of the above scenaria is that the enhancement of the curvature spectrum
by several orders of magnitude can be achieved only under special conditions.
The inflaton potential around an inflection point must be fine-tuned with high accuracy \cite{Garcia-Bellido:2017mdw, Germani:2017bcs, Motohashi:2017kbs, Dalianis:2018frf}.
A step in the potential gives a limited enhancement, unless the potential is 
engineered in a very specific way \cite{hu}. Finally, a sharp turn in the 
inflaton trajectory must reach a value of $4\pi$ for the enhanced spectrum to 
have observable consequences \cite{Palma:2020ejf}. This is possible only if the 
field manifold is curved.

Our aim is to demonstrate that a strong enhancement of the curvature spectrum  
is possible if several strong features in the inflaton potential combine 
constructively. For example, a sequence of steps or turns in field space, 
that occur within a small number of efoldings, can lead to an effect on the spectrum
that exceeds the effect of one feature by several orders of magnitude.

\section{Steps or inflection points} \label{notes}

In this section we analyze the scenario in which 
$\eta_\perp^2 \ll M^2/H^2$ during the whole evolution. The isocurvature mode
is always suppressed and the single relevant perturbation can be identified with
the curvature mode. Its evolution is governed by eq. (\ref{RfN}). 
This equation has been discussed in refs. \cite{Kefala:2020xsx, Dalianis:2021iig} in the
context of single-field inflation,
where oscillatory patterns have been observed.  In single-field setups oscillatory behaviour in specific parts of the spectrum has been also observed in ref. \cite{Ballesteros:2018wlw}.
 Our aim here is to obtain a better analytic
understanding of the nature of the solutions and the resulting form of the 
spectrum. We focus mainly on inflaton potentials that display one or more 
steps, which have been shown to result in power spectra of curvature perturbations
that can be strongly enhanced within a 
certain momentum range relative to the scale invariant case.

\subsection{Integral equations and an alternative formulation} \label{alternative} 

For a general form of the function $\eta_\parallel(N)$, 
we rewrite eq. (\ref{RfN}) as 
\be
R_{k,NN}+3R_{k,N}+\frac{k^2}{e^{2N} H^2} R_{k}=
2\eta_\parallel(N)R_{k,N}.
\label{ANg} \ee
The Green's function $G(N,n)$ for the operator in the lhs satisfies the equation 
\be
G_{k,NN}(N,n)+3G_{k,N}(N,n)+\frac{k^2}{e^{2N} H^2} G_{k}(N,n)=\delta(N-n).
\label{ANGr} \ee
The evolution is classical, so we must use the retarded Green's function, which 
satisfies $G_{k>}(N,n)=0$ for $n>N$. For $n<N$ the solution is 
\be
G_{k<}(N,n)= e^{-\frac{3}{2}\, N} \left(A(n) J_{3/2}\left( e^{- N}\frac{k}{H} \right) + B(n)\, J_{-3/2}\left( e^{-N}\frac{k}{H} \right)  \right).
\label{Grless} \ee
The Green's function is continuous at $N=n$. Its first derivative has a discontinuity, obtained
by integrating eq. (\ref{ANGr}) around $N=n$. This gives $G_{k<,N}(n,n)=1.$
Imposing these constraints gives
\begin{eqnarray}
A(n)&=&-\sqrt{\frac{\pi}{2}}e^{3n}\left(\frac{k}{H} \right)^{-3/2}
\left(\cos\left(e^{-n}\frac{k}{H} \right) +e^{-n}\frac{k}{H}\sin\left(e^{-n}\frac{k}{H} \right)\right)
\label{An} \\
B(n)&=&\sqrt{\frac{\pi}{2}}e^{3n}\left(\frac{k}{H} \right)^{-3/2}
\left(e^{-n}\frac{k}{H}\cos\left(e^{-n}\frac{k}{H} \right) -\sin\left(e^{-n}\frac{k}{H} \right)\right).
\label{Bn} \end{eqnarray}
Then, the solution of eq. (\ref{ANg}) is
\be
R_k(N)=\Rb_k(N;1,i,3)+2\int_{-\infty}^{N}G_k(N,n)\,
\eta_\parallel(n)\,R_{k,n}(n)\, dn,
\label{solg} \ee
with $\Rb_k(N;C_p,C_m,\kappa)$ given by 
\be
\Rb_k(N;C_{p},C_{m},\kappa)=A_0 \, e^{-\frac{1}{2}\kappa\, N} \left(C_{p}J_{\kappa/2}\left( e^{- N}\frac{k}{H} \right) + C_{m}\, J_{-\kappa/2} \left( e^{-N}\frac{k}{H} \right)  \right).
\label{solBessel} \ee
The values $\kappa=3$, $C_p=1$, $C_m=i$ correspond to the 
Bunch-Davies vacuum.

An approximate expression, which can be considered as the first step in an iterative solution of
the above equation, can be obtained if we replace the full solution $R_k(n)$ in the integral
with $\Rb_k(n,1,i,3)$. This gives \cite{Dalianis:2021iig} 
\begin{eqnarray}
&&R_k(\infty)=\Rb_k(\infty;1,i,3) \times
\nonumber \\
&&\,\left\{ 1-2i \frac{H}{k}
\int_{-\infty}^{\infty}\eta_\parallel(n)\,e^n 
\left[ e^{-n}\frac{k}{H}\cos\left(e^{-n}\frac{k}{H} \right) -\sin\left(e^{-n}\frac{k}{H} \right)\right]
\,
\left[ i \cos\left( e^{-n}\frac{k}{H} \right)-\sin\left( e^{-n}\frac{k}{H} \right) \right]
\, dn \right\}.
\nonumber \\
&~&
\label{solg2} \end{eqnarray}

Through a partial integration, we can recast eq. (\ref{solg}) as 
 \be
 R_k(N)=\Rb_k(N;1,i,3)-2\int_{-\infty}^{N}\frac{\partial}{\partial n}
 \left[ G_k(N,n)\,\eta_\parallel(n)\right]\,R_{k}(n)\, dn,
 \label{solgNpartial} \ee
 where we have used that $G_k(N,N)=0$ and $\eta_\parallel(n)\to 0$ for $n\to-\infty$.
The form (\ref{Grless}) of the Green's function suggests the ansatz
\be
R_k(N)= e^{-\frac{3}{2}\, N} \left(D(N) J_{3/2}\left( e^{- N}\frac{k}{H} \right) + E(N)\, J_{-3/2}\left( e^{-N}\frac{k}{H} \right)  \right).
\label{ansatz} \ee
Substituting in eq. (\ref{solgNpartial}) and matching the coefficients of the
Bessel functions, we obtain
\begin{eqnarray}
D(N)&=&1-\int_{-\infty}^N 
\frac{\partial}{\partial n} \left[ A(n)\,2\eta_\parallel(n)\right]\, e^{-\frac{3}{2}n}
\left[  J_{3/2}\left( e^{- n}\frac{k}{H} \right) D(n) 
+  J_{-3/2}\left( e^{-n}\frac{k}{H} \right) E(n) \right]dn
\label{DNint} \\
E(N)&=&i-\int_{-\infty}^N 
\frac{\partial}{\partial n} \left[ B(n)\,2\eta_\parallel(n) \right]\, e^{-\frac{3}{2}n}
\left[  J_{3/2}\left( e^{- n}\frac{k}{H} \right) D(n)
+ J_{-3/2}\left( e^{-n}\frac{k}{H} \right) E(n) \right]dn.
\label{ENint} \end{eqnarray}
In the first step of an iterative solution of the above equation, one 
substitutes $D(n)=1$ and $E(n)=i$ within the integrals in the rhs.
For $N\to \infty$ the solution (\ref{ansatz}) is dominated by the term proportional to $J_{-3/2}$.
Making use of a partial integration within the integral in eq. (\ref{ENint}),
we recover the approximate solution (\ref{solg2}).

In order to improve on this result, we can differentiate the relations (\ref{DNint}), (\ref{ENint})
with respect to $N$, in order to obtain a system of two first-order differential equations:
\begin{equation}
\frac{\partial}{\partial N}\begin{pmatrix} D(N)\\ E(N) \end{pmatrix}=
F(N)
\begin{pmatrix} D(N)\\ E(N) \end{pmatrix},
\label{system} \end{equation}
where
\be
F(N)=
-\begin{pmatrix} 
\frac{\partial}{\partial N}  \left[ A(N)\,2\eta_\parallel(N)\right]\, e^{-\frac{3}{2}N}
 J_{3/2}\left( e^{- N}\frac{k}{H} \right)&
~~~\frac{\partial}{\partial N} \left[ A(N)\,2\eta_\parallel(N)\right]\, e^{-\frac{3}{2}N}
 J_{-3/2}\left( e^{- N}\frac{k}{H} \right)
\\ 
\frac{\partial}{\partial N} \left[ B(N)\,2\eta_\parallel(N)\right]\, e^{-\frac{3}{2}N}
 J_{3/2}\left( e^{- N}\frac{k}{H} \right)&
~~~\frac{\partial}{\partial N} \left[ B(N)\,2\eta_\parallel(N)\right]\, e^{-\frac{3}{2}N}
 J_{-3/2}\left( e^{- N}\frac{k}{H} \right) \end{pmatrix}.
 \label{matr} \ee
The matrix $F(N)$ has vanishing determinant. Its nonzero eigenvalue is $2\eta_\parallel(N)$.
The system of equations (\ref{system}) can be solved numerically with the
initial condition $(D(N),E(N))=(1,i)$ that corresponds to the Bunch-Davies
vacuum. This formulation provides an advantage over the numerical solution
of eq. (\ref{RfN}), for which the initial condition is strongly oscillatory.
Moreover, it makes it straightforward to analyze alternative assumptions for
the vacuum. 

The solution of eq. (\ref{system}) can be expressed as  
\begin{equation}
\begin{pmatrix} D(N)\\ E(N) \end{pmatrix}=
Q(N)
\begin{pmatrix} 1\\ i \end{pmatrix},
\label{systemsol} \end{equation}
where $Q(N)$ is the fundamental matrix. 
An exact analytic determination of $Q(N)$ in closed form is not possible because of the $N$-dependence of
$F(N)$. For a slowly varying $F(N)$ (adiabatic limit), an approximate solution is 
given by 
\be
Q(N)=\exp(C(N)),~~~~~C(N)=\int_{-\infty}^N F(n)\, dn.
\label{CNint} \ee

In appendix \ref{appendixa} we provide an assessment of the 
accuracy of the approximate expressions (\ref{solg2}) and  (\ref{CNint}).
Both approximations give a very accurate
description of the spectrum when its value is of order 1. When
the spectrum is significantly enhanced both approximations lose accuracy.
However, eq. (\ref{CNint}) gives a reasonable approximation to the 
maximal value of the spectrum and its characteristic frequencies, even for an
enhancement by three orders of magnitude.

\begin{figure}[t!]
\centering
\includegraphics[width=0.6\textwidth]{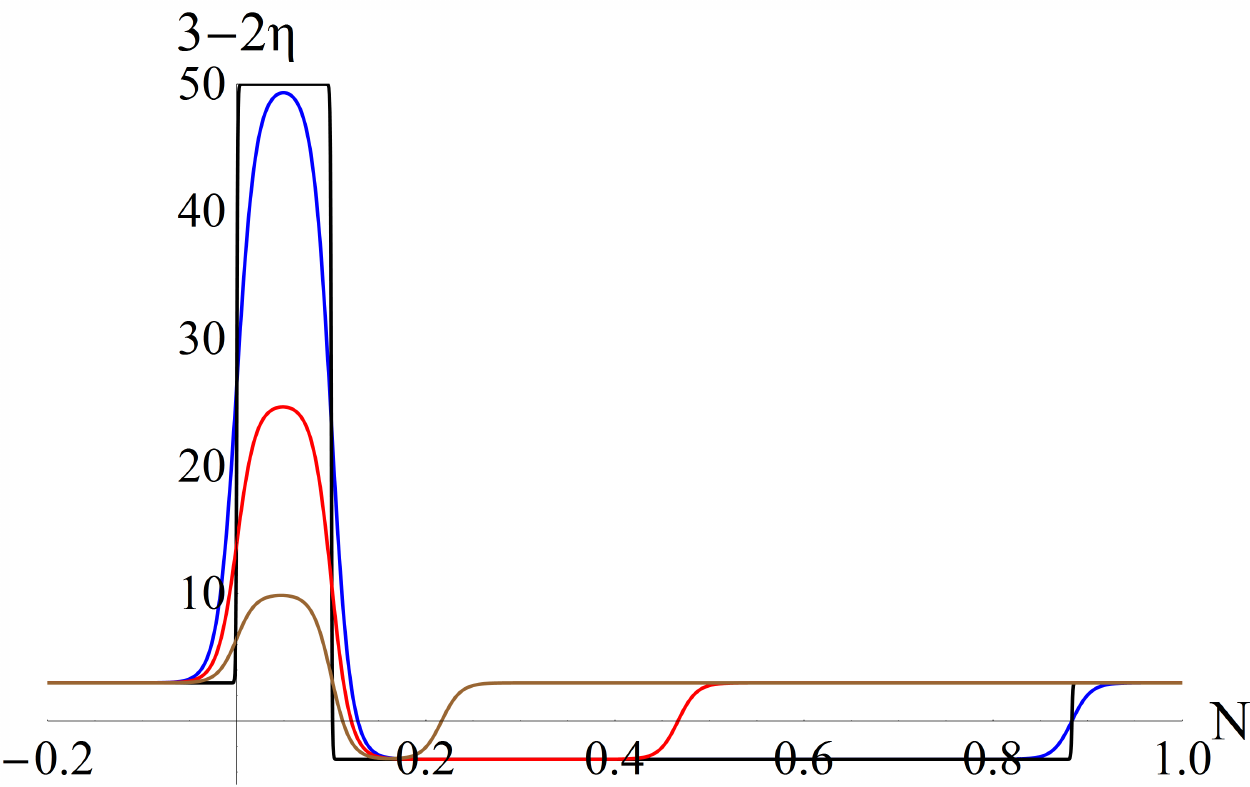} 
\caption{
Typical evolution of the function $\eta_\parallel(N)$ induced by steps in
the inflaton potential. 
}
\label{figzero}
\end{figure}

\subsection{Analytical expressions for ``pulses"} \label{analyticpulses}

In order to obtain a better understanding of the characteristic features
of the spectrum,
we consider a sequence of typical patterns for the function $\eta_\parallel(N)$, 
similar to those induced by steep
steps in the inflaton potential. The slow-roll parameter first decreases
sharply to very negative values during the very short interval that
the step is crossed by the background field.
Subsequently, the inflaton settles in a 
slow-roll regime on a flat part of the potential. During the approach to
slow-roll, the evolution is dominated by the first two terms in the 
equation of motion of the background field: $\ddot{\phi}+3H\dot{\phi}=0$.
This means that we have 
$\eta_\parallel\simeq 3$ during this interval. Eventually, the system returns 
to slow roll and 
$\eta_\parallel$ becomes almost zero.
It has been shown in refs. \cite{Kefala:2020xsx, Dalianis:2021iig} that 
the power spectrum is scale invariant  at late times (or $N\to \infty$) for 
$ k\rightarrow\infty$,  but has a value multiplied
by the factor \cite{Dalianis:2021iig}
\be \label{enhancement}
\left[ \delta \Delta^{(\infty)}_{R}\right]^2
=\exp\left(2\int_{-\infty}^\infty \eta_\parallel(N)  \, dN\right),
\ee 
relative to its scale-invariant value for modes that have sufficiently small $k$ and 
freeze before being affected by the features in the potential.
For $\eta_\parallel <0$ the spectrum is suppressed, while 
for $\eta_\parallel >0$ it is enhanced. 
If the integral is zero, the spectrum 
for $k\to \infty$ has the same amplitude
as for $k\to 0$. 
A different way to understand this effect is to notice that 
\be
-2\int_{N_e}^{N_l} \eta_{\parallel}(N)\,dN 
\simeq 2\log\frac{\epsilon_l}{\epsilon_e},
\label{integra} \ee 
where $\epsilon_e$ and $\epsilon_l$ are the values of the $\epsilon$ parameter
on the plateaus of the inflaton potential before and after a strong feature
that induces strong deviations from scale invariance. 
The integral vanishes 
for a feature localized between
regions supporting slow-roll inflation with similar values of $\epsilon$.
We impose this constraint on $\eta_\parallel(N)$ throughout this work
in order to focus solely on the effect 
of the feature.

In fig. \ref{figzero} we depict a sequence of smooth functions 
$\eta_\parallel(N)$ with
the above characteristics. The evolution is very similar to that
typically induced by inflaton potentials with steep steps 
\cite{Kefala:2020xsx, Dalianis:2021iig}.
The maximal value of $3-2\eta_\parallel$ increases with 
the steepness of the step, along with the duration of the interval
during which $3-2\eta_\parallel\simeq -3$.

In this subsection we present an analytic approach based on
an approximation of the evolution of $\eta_\parallel$ through ``pulses".
The essence of this approximation is depicted in fig. \ref{figzero} by the
curve with the very sharp transitions when $\eta_\parallel$ changes value. 
We name the evolution from vanishing $\eta_\parallel$ to a nonzero constant
value, and
back to zero after a period of efoldings, a square ``pulse".
For constant $\kappa \equiv 3-2\eta_{\parallel}$, 
the solution of eq. (\ref{RfN})
involves a linear combination of the Bessel functions 
$J_{\pm\kappa/2}$ and has the form of eq. (\ref{solBessel}).
For $\kappa=3$ we obtain a scale-invariant spectrum. 
The Bunch-Davies vacuum corresponds to
$C_p=1$, $C_m=i$. As we have mentioned before, 
we are interested only in relatively strong deviations from scale invariance, 
and not
in the absolute normalization of the spectrum. For this reason we set $A_0=1$.

If $\kappa$ changes at $N=N_{fi}$ from an initial value $\kappa_{i}$ 
to a new value $\kappa_{f}$, one can 
compute the change from initial coefficients $C_{p_{i}}, C_{m_{i}}$ to 
new coefficients $ C_{p_{f}}, C_{m_{f}}$ by requiring the 
continuity of the solution of eq. (\ref{RfN}) and
its first derivative \cite{Dalianis:2021iig}.
The new coefficients  are given by
\be
\begin{pmatrix}  
		 C_{p_{f}}\\
		  C_{m_{f}}
\end{pmatrix}=M(N_{fi},\kappa_{i},\kappa_{f},k)
\begin{pmatrix}	  
		 C_{p_{i}}\\
		  C_{m_{i}}
\end{pmatrix},
\label{coefficients}	  
\ee
where the matrix $M(N_{fi},\kappa_{i},\kappa_{f},k)$
has components
\begin{eqnarray}
	M_{11}&=&C \left( J_{-{\kappa_{f}}/{2}}\left(e^{-N_{fi}}\frac{k}{H}\right)
	J_{-1+\kappa_{i}/2}\left(e^{-N_{fi}}\frac{k}{H}\right)
	+J_{1-{\kappa_{f}}/{2}}\left(e^{-N_{fi}}\frac{k}{H}\right)
	J_{{\kappa_{i}}/{2}}\left(e^{-N_{fi}}\frac{k}{H}\right)\right)
	\nonumber \\ 
	M_{12}&=&C \left( -J_{-{\kappa_{f}}/{2}}\left(e^{-N_{fi}}\frac{k}{H}\right)
	J_{1-{\kappa_{i}}/{2}}\left(e^{-N_{fi}}\frac{k}{H}\right)
	+J_{1-{\kappa_{f}}/{2}}\left(e^{-N_{fi}}\frac{k}{H}\right)
	J_{-{\kappa_{i}}/{2}}\left(e^{-N_{fi}}\frac{k}{H}\right)\right)
	\nonumber \\ 
	M_{21}&=&C\left( -J_{{\kappa_{f}}/{2}}\left(e^{-N_{fi}}\frac{k}{H}\right)
	J_{-1+\kappa_{i}/2}\left(e^{-N_{fi}}\frac{k}{H}\right)
	+J_{-1+\kappa_{f}/2}\left(e^{-N_{fi}}\frac{k}{H}\right)
	J_{{\kappa_{i}}/{2}}\left(e^{-N_{fi}}\frac{k}{H}\right)\right)
	\nonumber \\ 
	M_{22}&=&C\left( J_{{\kappa_{f}}/{2}}\left(e^{-N_{fi}}\frac{k}{H}\right)
	J_{1-{\kappa_{i}}/{2}}\left(e^{-N_{fi}}\frac{k}{H}\right)
	+J_{-1+\kappa_{f}/2}\left(e^{-N_{fi}}\frac{k}{H}\right)
	J_{-{\kappa_{i}}/{2}}\left(e^{-N_{fi}}\frac{k}{H}\right)\right),
	\nonumber \\ &&
\label{Mab} \end{eqnarray}
with
\be
C=\frac{\pi}{2}e^{\frac{1}{2}N_{fi}(-2+\kappa_{f}-\kappa_{i})}\frac{k}{H} \,
\csc\left(\frac{\pi \kappa_{f}}{2}\right).
\label{Mconst} \ee
The matrix has the property $ M(N_{fi},\kappa_{m},\kappa_{f},k)\cdot M(N_{fi},\kappa_{i},\kappa_{m},k)=M(N_{fi},\kappa_{i},\kappa_{f},k)$, so
that we can select the value 
$\kappa=3$ as a reference point for all transitions between
different values of $\kappa$.

One can also define a matrix corresponding to a ``pulse" of height $\kappa$
different from the value 3 corresponding to the scale-invariant case. 
The parameter $\eta_\parallel$ is 
\be
\eta_\parallel(N)= \frac{3-\kx}{2} \left(\Theta(N-N_1)-\Theta(N-N_2) \right)
\label{etapartheta} \ee
 and the corresponding matrix has the form
\be
M_{\rm pulse}(N_1,N_2,\kappa,k)   =   M(N_{2},\kappa,3,k) \cdot M(N_{1},3,\kappa,k).
\label{Mpulse} \ee
A product of several $M_{\rm pulse}$ 
matrices can reproduce the final values of the coefficients 
$(C_p,C_m)$ of the Bessel functions $J_{\pm 3/2}$ after the perturbations have evolved
from an initial configuration corresponding to $(C_p,C_m)=(1,i)$ through a period 
of strong features. Clearly, it is possible to reconstruct
any smooth function $\eta_\parallel(N)$ in terms of 
short intervals of $N$ during which the function takes constant values. Multiplying the
corresponding $M_{\rm pulse}$ matrices would provide a solution to the problem of the 
evolution of perturbations. 
The increase of the power spectrum relative to the 
scale invariant one is given by the final value of the coefficient 
$|C_m|^2$ after a mode of given $k$
has evolved past the strong features in the background. 

Simple analytic expressions can be obtained in the limits of large and small $k$,
using the corresponding expansions of the Bessel functions.
Keeping the leading contribution, one finds that
the power spectrum is scale invariant  at late times (or $N\to \infty$) for $ k\rightarrow\infty$,  but has a value multiplied 
by the factor of eq. (\ref{enhancement}).
The corrections subleading in $H/k$ introduce oscillatory patterns in the spectrum
\cite{Dalianis:2021iig}. 

Explicit expressions can be obtained for the case of a ``pulse" of large positive
amplitude (negative $\eta_\parallel$) and very short duration. 
For large positive values of $\kappa$ and 
a short interval $\delta N$, 
we can approximate the Bessel function as 
\be
J_{a}(z)= \frac{1}{\Gamma(a+1)}
\left(\frac{z}{2} \right)^{a} \left(1-\frac{z^2}{4(1+a)}+{\cal O}(z^4) \right).
\label{expansion} \ee
We find that 
\be
M^{\rm (sharp)}_{\rm pulse}(N_1,k)=e^{N_1} \frac{H}{k}
\begin{pmatrix} T_{11}&T_{12}\\T_{21}&T_{22}
 \end{pmatrix} 
 \left(1+{\cal O}\left(\frac{1}{2\kappa}\left( e^{-N_1}\frac{k}{H} \right)^2 
 \right) \right),
\label{sharppulse} \ee
where
\begin{eqnarray}
T_{11}&=& \cos\left( e^{-N_1} \frac{k}{H}\right) \left( 
e^{-N_1} \frac{k}{H} \cos\left( e^{-N_1} \frac{k}{H}\right) 
-\sin\left( e^{-N_1} \frac{k}{H}\right)  \right), \nonumber \\
T_{12}&=&\cos\left( e^{-N_1} \frac{k}{H}\right) \left( 
e^{-N_1} \frac{k}{H} \sin\left( e^{-N_1} \frac{k}{H}\right) 
+\cos\left( e^{-N_1} \frac{k}{H}\right)  \right),
\nonumber \\
T_{21}&=&\sin\left( e^{-N_1} \frac{k}{H}\right) \left( 
e^{-N_1} \frac{k}{H} \cos\left( e^{-N_1} \frac{k}{H}\right) 
-\sin\left( e^{-N_1} \frac{k}{H}\right)  \right),
\nonumber \\
T_{22}&=&\sin\left( e^{-N_1} \frac{k}{H}\right) \left( 
e^{-N_1} \frac{k}{H} \sin\left( e^{-N_1} \frac{k}{H}\right) 
+\cos\left( e^{-N_1} \frac{k}{H}\right)  \right).
\label{Mpulsesharp} \end{eqnarray}

For $ k/H \lta e^{N_1} \sqrt{2\kappa}$ the presence of the ``pulse" does not
induce a suppression of the perturbation. The only effect is the introduction of
oscillations in the spectrum. 
The spectrum is expected to have a deep minimum once per period, i.e. 
at intervals $\delta k/H=e^{N_1}\pi$. The suppression of the 
pertubation by $\exp(-\kappa \,\delta N) $ is expected to occur for
$k/H \gta e^{N_1} \sqrt{2\kappa}$.

After a step in the potential is crossed, the inflaton settles in a 
slow-roll regime on a flat part of the potential. During the approach to
slow-roll, the evolution is dominated by the first two terms in the 
equation of motion of the background field: $\ddot{\phi}+3H\dot{\phi}=0$.
This means that, during an interval $N_1\leq N \leq N_2$, we have 
$\eta_\parallel\simeq 3$ and $\kappa=-3$.  
We can define the matrix 
\be
M^{(\rm negative)}_{\rm pulse}(N_1,N_2,k)  =  M(N_{2},-3,3,k) \cdot M(N_{1},3,-3,k)
\label{Mpulseneg} \ee
in order to account for the effect on the curvature perturbation, similarly to
the treatment above. 
Then, the total effect on the coefficients of the 
Bessel functions, arising from crossing a step in the potential, is given 
by
\be
\begin{pmatrix}   C_{p_{f}}\\ C_{m_{f}} \end{pmatrix}=
M^{(\rm negative)}_{\rm pulse}(N_1,N_2,k)\cdot  M^{(\rm sharp )}_{\rm pulse}(N_1,k) 
\begin{pmatrix}	   1 \\ i \end{pmatrix}.
\label{totaleff} \ee
The value of the spectrum relative to the scale-invariant case 
is determined by $|C_{m_f}|^2$.  
We can obtain an explicit expression for the enhancement of the spectrum
by evaluating $C_{m_f}$ through eq. (\ref{totaleff}), 
keeping the leading contribution for $M^{(\rm sharp )}_{\rm pulse}(N_1,k)$.
By defining $\kt=e^{-N_1}k/H$ and $\ex=\exp(-N_2+N_1)$, we find
\be
|C_{m_f}|^2=\frac{1+\kt^2}{4\ex^{12} \kt^{12}}
\left(A_1 \sin(\kt)+A_2\cos(\kt)+A_3 \sin(\kt-2\ex \kt) +A_4 \cos(\kt-2\ex \kt)\right)^2,
\label{coeffneg} \ee
with 
\begin{eqnarray}
A_1&=&9\kt+9\ex^2 \kt^3-6\ex^3\kt^3+2\ex^3\kt^5
\nonumber \\
A_2&=&9-3\kt^2+9\ex^2\kt^2-3\ex^2\kt^4+6\ex^3\kt^4
\nonumber \\
A_3&=&-9\kt+18\ex \kt-6\ex\kt^3+9\ex^2\kt^3
\nonumber \\
A_4&=&-9+3\kt^2-18\ex\kt^2+9\ex^2\kt^2-3\ex^2\kt^4.
\label{Acoeff} \end{eqnarray}

In order for a step to increase the spectrum by several orders of magnitude, one
must have $N_2-N_1\gta 1$. 
Modes with $e^{-N_1}k/H = {\cal O}(1)$ satisfy $e^{-N_2}k/H \ll 1$.
In this momentum range we can expand eq. (\ref{coeffneg}) in $\ex$ and keep the 
leading contribution
\be
|C_{m_f}|^2\simeq e^{6(N_2-N_1)}\frac{4(1+\kt^2)}{\kt^{6}}
\left(3\kt \cos(\kt)+(-3+\kt^2)\sin(\kt) \right)^2.
\label{coeffnegappr} \ee
The enhancement is given by eq. (\ref{enhancement}),
taking into account only the contribution from the negative ``pulse" (positive $\eta_\parallel$).
There is also oscillatory behaviour induced by sines and cosines of 
$\exp(-N_1)k/H$. The combined effect indicates that the 
power spectrum near its maximum is enhanced through the negative ``pulse", but
also develops strong oscillations at intervals $\delta k/H=e^{N_1}\pi$.
We emphasize that the expressions (\ref{coeffneg}) and (\ref{coeffnegappr}) 
are valid only near the maximum of the spectrum.
They do not account for the expected drop of the spectrum for
$ e^{-N_1}k/H \gta \sqrt{2\kappa}$, with $\kappa$ the height of the positive 
sharp ``pulse".

\begin{figure}[t!]
\centering
\includegraphics[width=0.48\textwidth]{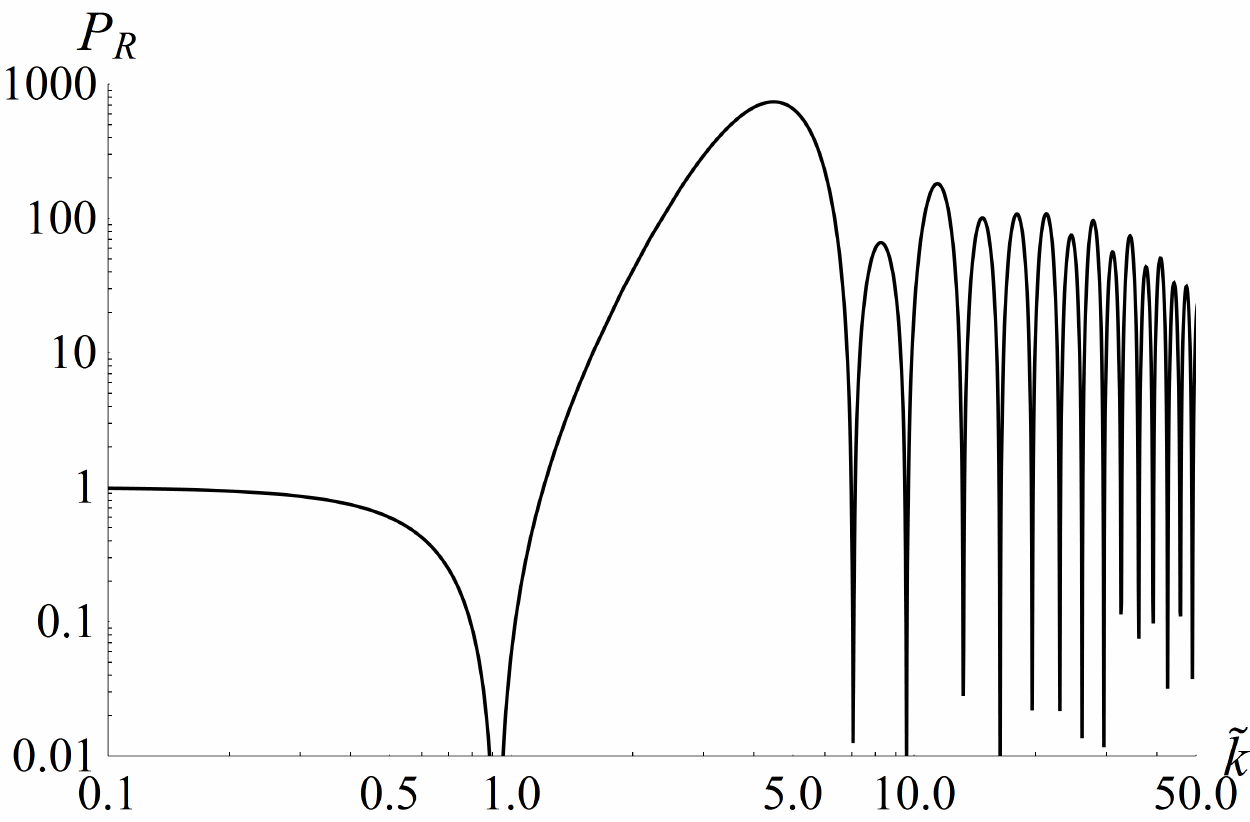} \includegraphics[width=0.48\textwidth]{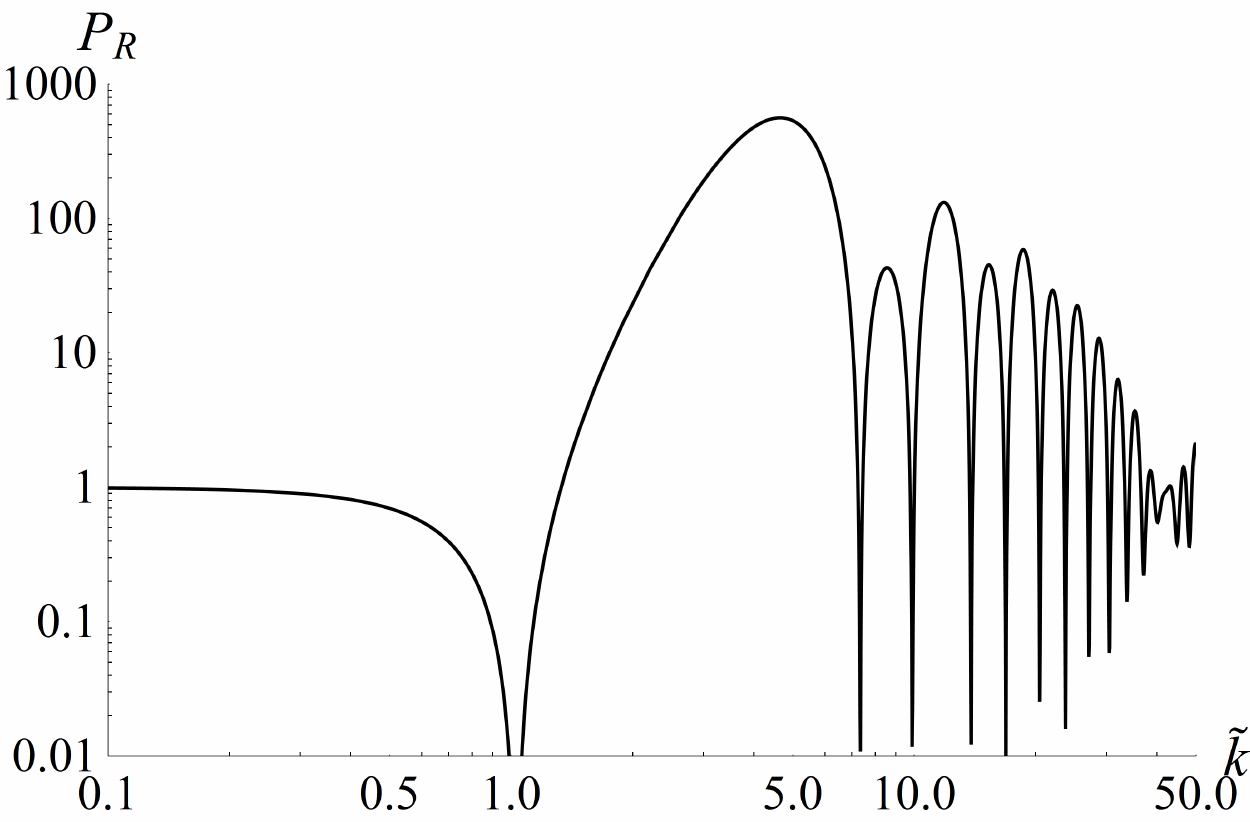} \includegraphics[width=0.48\textwidth]{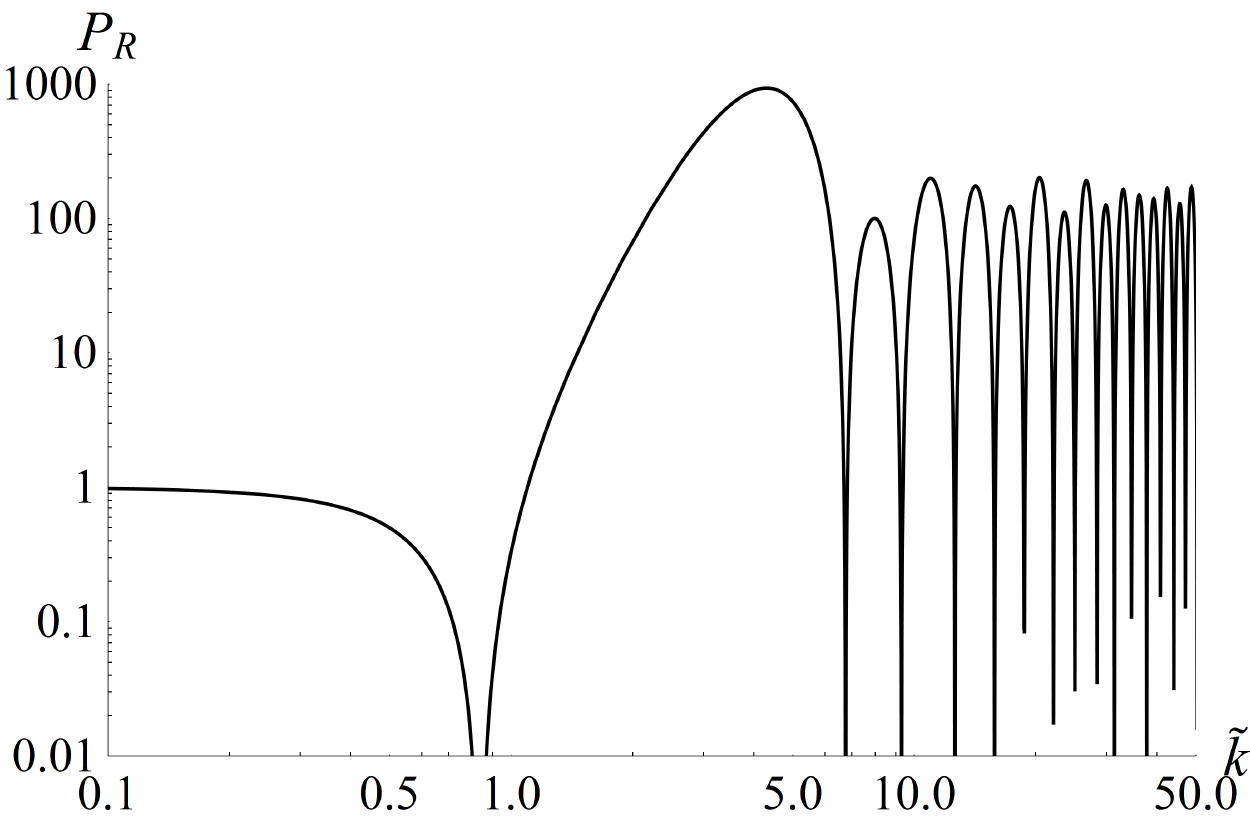} \includegraphics[width=0.48\textwidth]{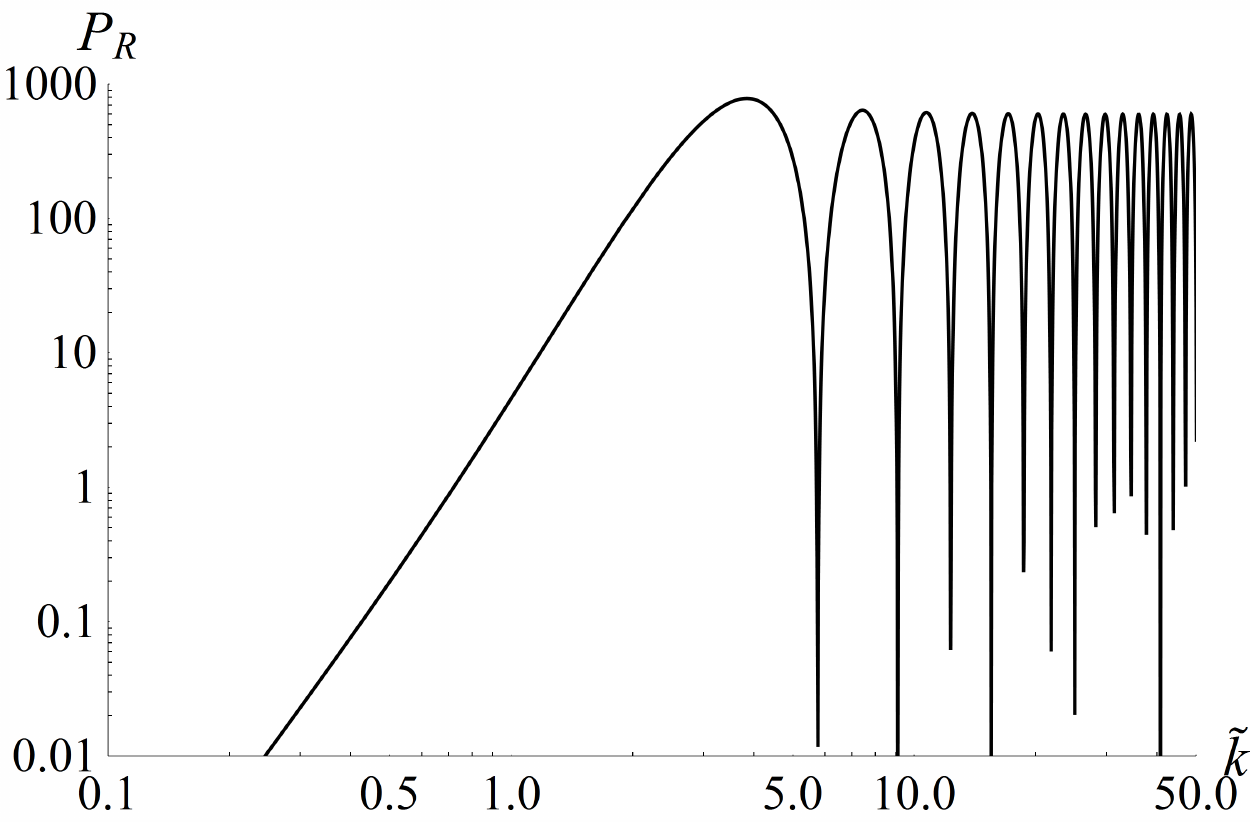}
\caption{
The power spectrum induced by a positive ``pulse" of height $\kappa$ between
$N_1$ and $N'_1$, followed by a negative ``pulse" of height $-3$ 
between $N'_1$ and $N_2$, as a function of $\kt=e^{-N_1}k/H$.
Top row, left plot: $\kappa=100$, $N_1=0$, $N'_1=0.05$, $N_2=0.858$.
Top row, right plot: $\kappa=50$, $N_1=0$, $N'_1=0.1$, $N_2=0.883$.
Bottom row, left plot: The approximation of eq. (\ref{coeffneg}) 
for $N_1=0$, $N_2=0.833$.
Bottom row, right plot: The approximation of eq. (\ref{coeffnegappr}) 
for $N_1=0$, $N_2=0.833$.
}
\label{figone}
\end{figure}

In fig. \ref{figone} we test the accuracy of the various analytic
results of this section in the case of square ``pulses", such as those
depicted in fig. \ref{figzero}. The top row displays the spectra for
two cases:
a) a positive ``pulse" of height $\kappa=100$ between
$N_1=0$ and $N'_1=0.05$, followed by a negative ``pulse" of height $-3$ 
between $N'_1=0.05$ and $N_2=0.858$, and 
b) a positive ``pulse" of height $\kappa=50$ between
$N_1=0$ and $N'_1=0.1$, followed by a negative ``pulse" of height $-3$ 
between $N'_1=0.1$ and $N_2=0.883$. 
These are simplified versions of the typical evolution of $\eta_\parallel$
when the background inflaton crosses steps in the potential of 
variable steepness \cite{Kefala:2020xsx, Dalianis:2021iig}.  
The integrated area of the negative ``pulse" is approximately the same in both cases, so
that a comparable enhancement of the spectrum is expected.
The spectra in the top row have been computed through the
numerical solution of eq. (\ref{RfN}), the system of eq. (\ref{system}) and
the expression of eq. (\ref{totaleff}). All these methods agree very well.
The bottom row displays the approximation of eq. (\ref{coeffneg})
(left plot) and the approximation of eq. (\ref{coeffnegappr}) (right plot).
Both these expressions assume $N_1'\simeq N_1$, while they 
do not depend on the height $\kappa$ of the positive ``pulse", as long
as this is very large. We have used $N_1=0$, $N_2=0.833$.
It is apparent that eq. (\ref{coeffneg}) reproduces very well the form of
the spectrum for $ e^{-N_1}k/H \lta \sqrt{2\kappa}$. For larger 
values of $k$, the expected drop of the spectrum is not captured by
this approximation. The deviation is clearer in the right plot, for
which $\kappa$ is smaller and the drop sets in earlier.
On the other hand, eq. (\ref{coeffnegappr}) is a cruder approximation.
However, its simplicity makes it very useful for estimating the
magnitude of the enhancement of the spectrum, as well as the 
fundamental frequency of oscillations.

\subsection{Multiple features}

In this subsection we 
examine the effect on the power spectrum of several ``pulses" in the evolution of 
$\eta_\parallel(N)$. A similar effect has already been 
discussed in refs. \cite{Kefala:2020xsx, Dalianis:2021iig} through a numerical solution of
eq. (\ref{RfN}). We return to this issue here, making use of the 
approximation of eq. (\ref{coefficients}). 
A similar effect will be discussed in the following section, arising through
$\eta_\perp$.

We specialize in the case of ``pulses" generated through steps in the 
inflaton potential. The effect of one step is captured by the
relation (\ref{totaleff}). 
In most cases one step is not sufficient to 
induce an enhancement of the spectrum by more than three or four orders of 
magnitude.
It must be noted, however, that an enhancement by up to seven orders of 
magnitude is possible 
for an appropriately engineered potential and step profile \cite{hu}.
The generalization to several steps is straightforward,
through the inclusion of several $M_{\rm pulse}$ matrices, and can 
lead to further enhancement.

\begin{figure}[t!]
\centering
\includegraphics[width=0.48\textwidth]{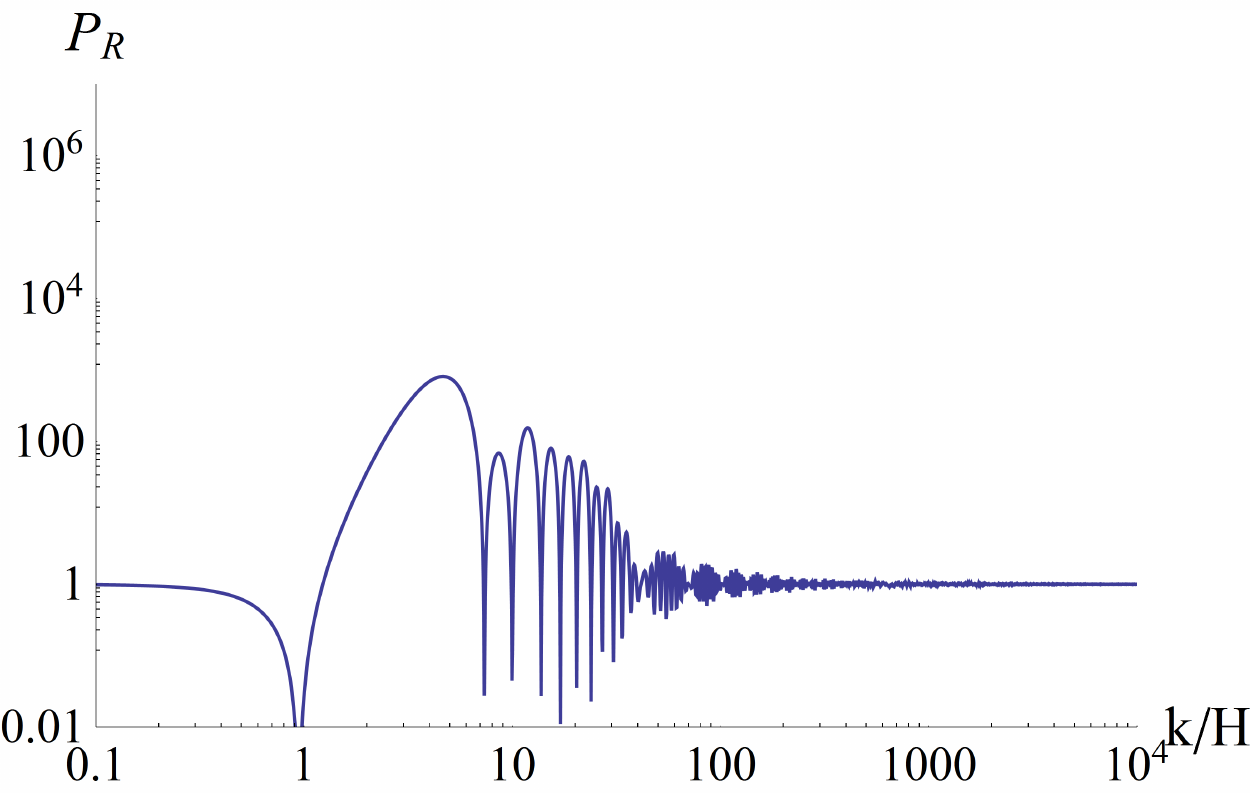} \includegraphics[width=0.48\textwidth]{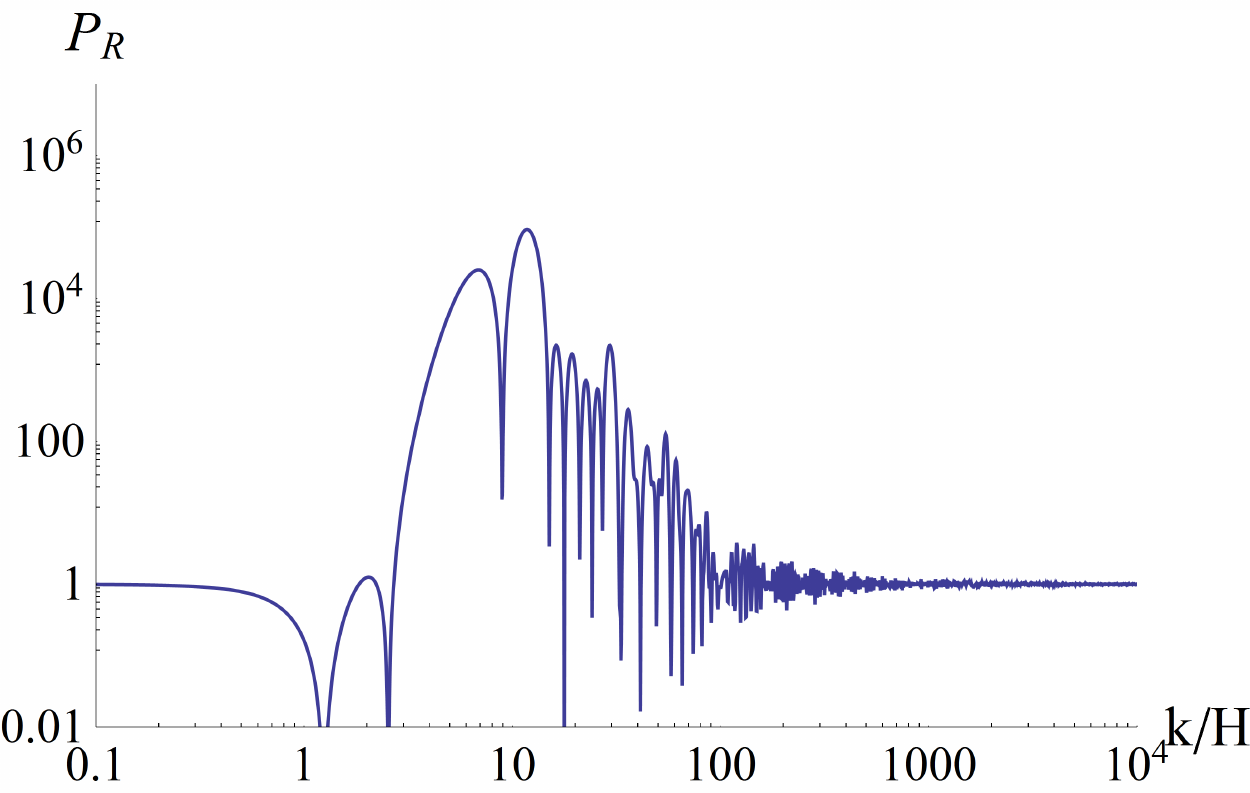} \includegraphics[width=0.48\textwidth]{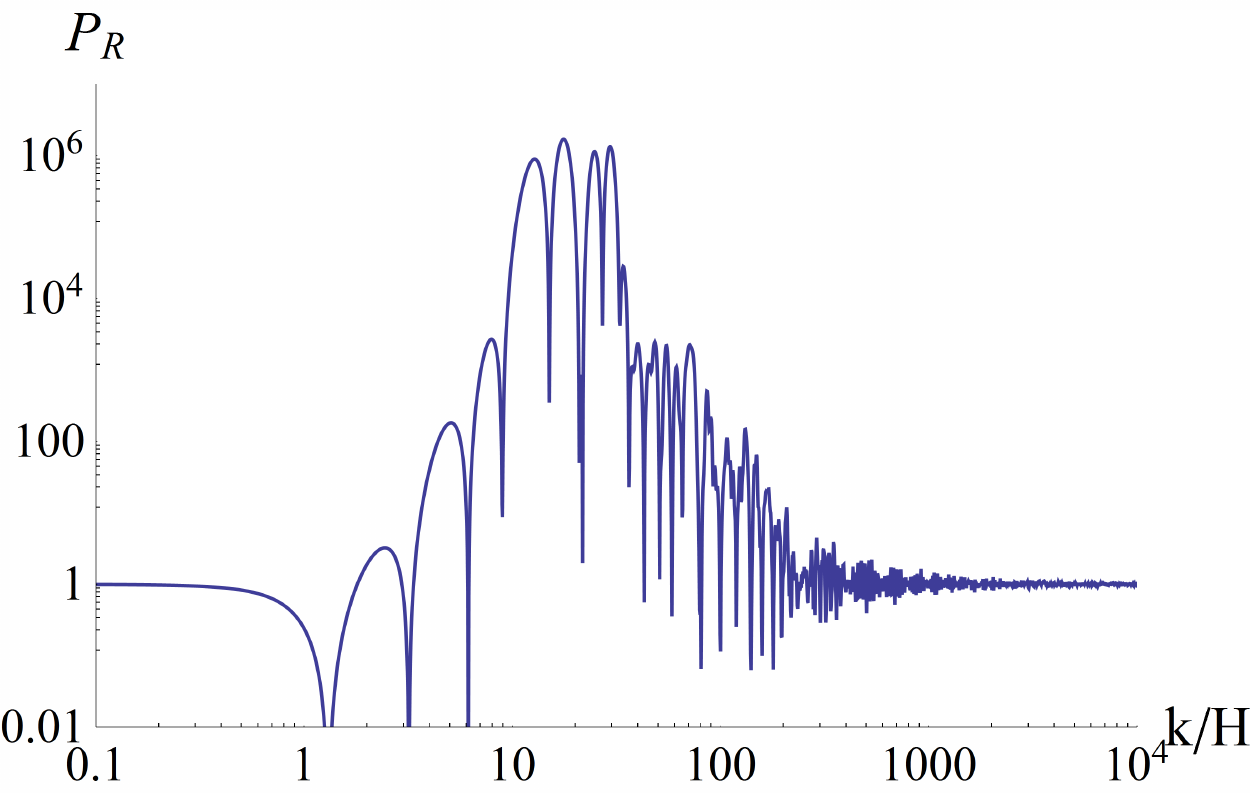} \includegraphics[width=0.48\textwidth]{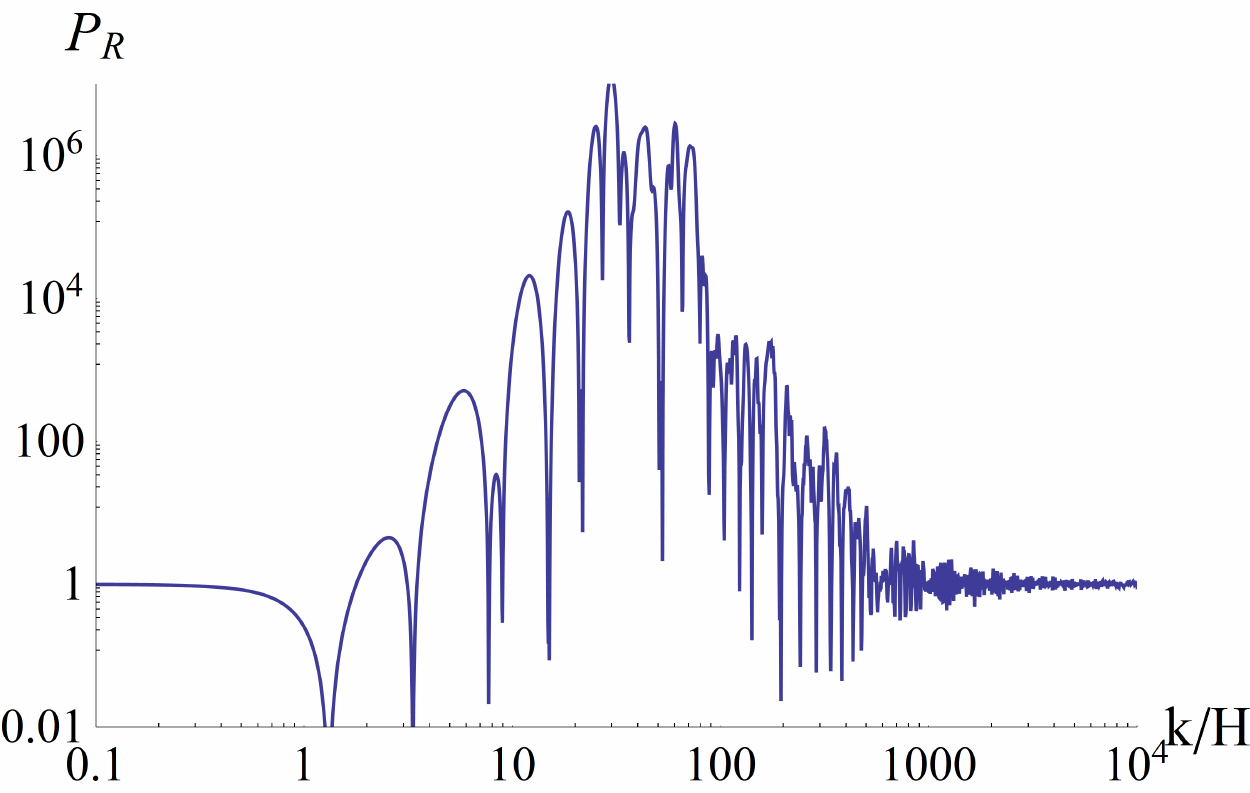}
\caption{
Power spectra induced by successive occurrences of a feature consisting of 
a positive ``pulse" of height $\kappa=50$ between
$N_1=0$ and $N'_1=0.1$, followed by a negative ``pulse" of height $\kappa=-3$ 
between $N'_1=0.1$ and $N_2=0.883$.
Top row: one feature (left plot), two features (right plot).
Bottom row: three features (left plot), four features (right plot).
}
\label{figtwo}
\end{figure}

If fig. \ref{figtwo} we present spectra induced by one or more features in
the evolution of $\eta_\parallel$, consisting of 
a positive ``pulse" of height $\kappa=50$ between
$N_1=0$ and $N'_1=0.1$, followed by a negative ``pulse" of height $\kappa=-3$ 
between $N'_1=0.1$ and $N_2=0.883$. We present four spectra, arising when
one, two (top row), three or four (bottom row) such features occur, one immediately
after the other. 
As has been discussed in refs. \cite{Kefala:2020xsx, Dalianis:2021iig}, such features
result from multiple steep steps in the inflaton potential.
The spectra obtained for one feature are consistent with 
those derived through similar approaches \cite{Ozsoy:2019lyy, Tasinato:2020vdk}.
Moreover, it is apparent that multiple features act constructively, 
increasing the enhancement of the spectrum.
A multitude of characteristic frequencies also appear through the 
dependence of the spectrum on 
combinations of the form $e^{-N_i}k/H$, with $N_i$ corresponding 
to the time that a sharp transition occurs in the evolution of $\eta_\parallel (N)$.

It is important to note that the various sets of 
``pulses" should be placed close to each other 
for the enhancement of the power spectrum to be significant. Similarly,
the corresponding
steps in the inflaton potential must be close.
If we increase the distance in efoldings between the sets of ``pulses", 
the additive effect is not as strong. Beyond a certain distance, 
each set acts independently on the spectrum, giving a moderate enhancement
within a different $k$-range.

\section{Turns in field space}\label{sectionturns}

In this section we turn to a different source of enhancement of the 
curvature spectrum, which is possible when
the isocurvature perturbations become strongly excited during 
a finite period of efoldings and act as a source for the curvature 
perturbations. In the two-field case we discussed in section 
\ref{twofieldsection}, this situation occurs when the parameter 
$\eta_\perp$ satisfies $\eta_\perp^2 \gg M^2/H^2$, with $M$ the typical
mass of the mode perpendicular to the inflaton trajectory. The relevant
equations are eqs. (\ref{RRR}), (\ref{SSS}). The evolution of the background
can be rather complicated, depending on the characteristics of the 
two-field potential \cite{Cespedes:2012hu, Achucarro:2012yr}. We concentrate here on
a simplified scenario, which preserves the relevant features 
without requiring a numerical calculation of the background evolution.

\subsection{Maximal turn and multiple features} \label{maximal}

We consider models of a two-component 
field $\phi^a=(\chi,\psi)$ with a standard kinetic term, for which 
the curvature of the field manifold vanishes.
We assume that the potential has an almost flat direction along a curve
$\psi=f(\chi)$. A small constant slope along this direction results in 
a small value of the slow-roll parameter $\epsilon$, which we assume to be constant. 
Along the perpendicular direction the potential has 
a large curvature, so that the flat direction forms a valley. 
We consider the simplest scenario, in which the fields evolve very close
to the bottom of the valley, without perpendicular oscillations.
We approximate the Hubble parameter as constant, with a value set by the average
value of the potential along the flat direction. A particular realization of
this setup, with $f(\chi)=a^2/\chi$, is given in ref. \cite{Cespedes:2012hu}.

The unit vectors, tangential and normal to the valley at
$\phi^a=\left(\chi, f(\chi) \right)$, are
\begin{eqnarray}
T^a&=&\frac{1}{\sqrt{1+f'^2(\chi)}} \left(1, f'(\chi) \right)
\label{tangunit} \\
N^a&=&\frac{1}{\sqrt{1+f'^2(\chi)}} \left(f'(\chi),-1 \right).
\label{tangunit} \end{eqnarray}
If we assume that the norm of $\dot{\phi}^a$ stays constant and equal 
to $\sqrt{2\epsilon}\,H$,
we find that 
\be
\chi_{,N}=\pm \frac{ \sqrt{2\epsilon}}{\sqrt{1+f'^2(\chi)}}. 
\label{dchidt} \ee 
We now have
\be
T^a_{,N}=\frac{\partial T^a}{\partial \chi}\chi_{,N}=
\frac{f''(\chi)}{(1+f'^2(\chi))^{3/2}}\, \chi_{,N}\,(-f'(\chi),1)=
-\frac{f''(\chi)}{1+f'^2(\chi)}\, \chi_{,N}\, N^a.
\label{eq1} \ee
From eq. (\ref{Neta})
we deduce that
\be
\eta_\perp  = \frac{f''(\chi)}{1+f'^2(\chi)}\chi_{,N}
=\pm \sqrt{2\epsilon} \frac{f''(\chi)}{(1+f'^2(\chi))^{3/2}}.
\label{etaperp} \ee
It is apparent that $\eta_\perp$ is nonzero only in regions in which
$f''(\chi)\not=0$, so that the valley of the potential is not linear. 

We are interested in a scenario in which a linear part of the valley is
succeeded by a sharp turn, leading to a second linear part. Without loss
of generality we can assume that the turn is located near $\chi=\psi=0$. 
The background
evolution will be characterized by a short interval in which $\eta_\perp$ will
rise and fall sharply from zero.
The angle of rotation in field space obeys
$\theta_{,N}= \eta_\perp,$
from which we obtain
\be
 \Delta \theta =\int_{N_i}^{N_f} \eta_\perp(\chi) dN
=\int_{\chi_i}^{\chi_f} \frac{f''(\chi)}{1+f'^2(\chi)} d\chi=
\Bigl. \arctan\left( f'(\chi) \right) \Bigr|_{\chi_i}^{\chi_f}.
\label{thetatotal} \ee
The maximal angle can be obtained if $f'(\chi_i)\to-\infty$ before the 
turn and $f'(\chi_f)\to \infty$ after, so that $\Delta\theta=\pi$.

In the model of ref. \cite{Cespedes:2012hu}, in which $f(\chi)=a^2/\chi$, 
one can have $f'(\chi_i)\to -\infty$ before the 
turn and $f'(\chi_f)\to 0$ after, so that $\Delta\theta=\pi/2$.
The maximal value of $\eta_\perp $ is obtained for $\chi=\psi=a$. It is 
$\eta_{\perp{\rm max}}={\sqrt{\epsilon}}/{a}$, and can be arbitrarily
large for $a\to 0$. The duration of the turn is roughly 
$\Delta N\sim \Delta\theta /\eta_{\perp{\rm max}}=\pi a/(2\sqrt{\ex})$
and can be very short for $a\to 0$.

\begin{figure}[t!]
\centering
\includegraphics[width=0.6\textwidth]{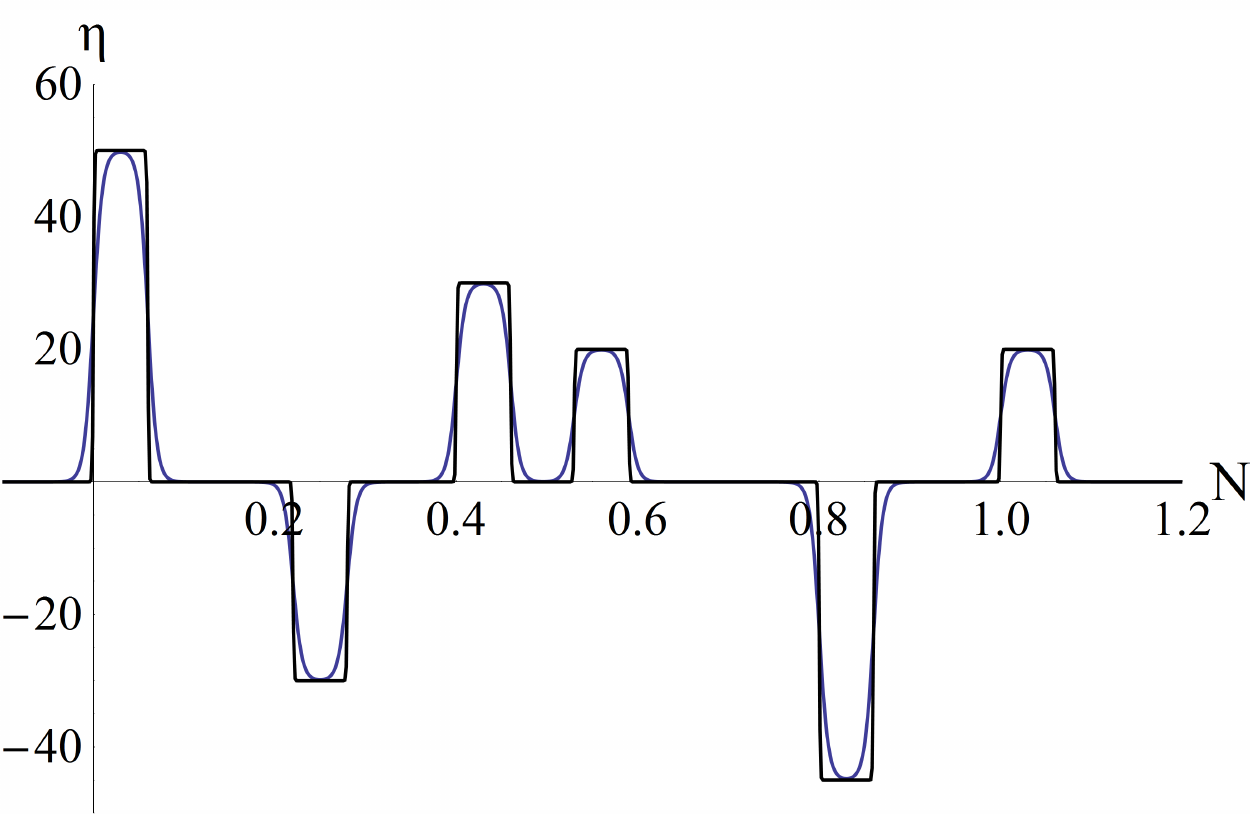} 
\caption{
Typical evolution of the function $\eta_\perp(N)$ induced by turns in
the inflaton potential. The approximation through ``pulses" is also depicted.
}
\label{etaperpp}
\end{figure}

As the value of the integral in eq. (\ref{thetatotal})
is bounded by $\pi$, the effect of sharp turns on the amplification of 
the isocurvature and curvature perturbations is limited. However, multiple
turns can also occur. The sign of the rotation angle is arbitrary, so 
a sequence of turns with alternating signs is possible. 
The enhancement of the curvature perturbation depends only on $|\eta_\perp|$,
as can be easily seen through inspection of eqs. (\ref{RRR}), (\ref{SSS}).
In fig. \ref{etaperpp} we depict the typical evolution of $\eta_\perp$ when
the potential has several turns along its flat direction.
We also present the approximation of the various features through ``pulses", 
which we shall employ in the following. The integral over each
feature must be smaller than $\pi$, so that the valley of the potential does 
not close on itself. This limits the possible enhancement arising from each
turn. However, the combined effect of several turns can be substantial, as we
show in this section.

\subsection{The qualitative features of the evolution} \label{qualitative}

As we discussed in section \ref{twofieldsection} the perturbations in the
two-field system are governed by eqs. (\ref{RRR}). (\ref{SSS}).
In order to simplify the picture, we assume that $M$ is constant and 
$M/H\gta 1$, so that the isocurvature modes get suppressed
during the parts of the evolution in which $\eta_\perp$ is small. 
However, when $\eta_\perp \gg M/H$ the evolution becomes
non-trivial, as both the curvature and isocurvature modes
get excited. The term $\sim \eta^2_\perp$ in the lhs of 
eq. (\ref{SSS}) acts as a negative mass term, triggering the 
rapid growth of the field $\Sc_k$. As a result, the rhs of eq. (\ref{RRR})
becomes a strong source term for the field $\Rc_k$. The resulting growth 
of $\Rc_k$ generates a source term in the rhs of eq. (\ref{SSS}) that
moderates the maximal value of the field $\Sc_k$. 
The combined effect results in the enhancement of both modes. However,
the late evolution of $\Sc_k$, when $\eta_\perp$ becomes negligible, is dominated 
by its nonzero mass, so that this field eventually vanishes. The curvature
mode $\Rc_k$ freezes, similarly to the standard inflationary scenario, preserving its enhancement within a certain momentum range.
Another characteristic consequence of the presence of strong 
features in the potential during inflation, is the appearance of 
distinctive oscillations in the curvature spectrum.

In order to obtain an analytic understanding of the evolution of the perturbations, we
consider the ``pulse" approximation of the previous section, in which 
$\eta_\perp$ has the form
\be
\eta_\perp(N)=\etapz\left( \Theta(N-N_1)-\Theta(N-N_2)\right).
\label{etaperppulse} \ee
The 
parameter $\eta_\perp$ takes a constant value for $N_1<N<N_2$, and approaches zero very
quickly outside this range.
For $N<N_1$ and $N>N_2$ 
the two equations of motion decouple
\begin{eqnarray}
\mathcal{R}_{k,NN}+3\mathcal{R}_{k,N}+\frac{k^2}{H^2}e^{-2N}\mathcal{R}_k&=&0
\label{eqmot1} \\
\mathcal{S}_{k,NN}+3\mathcal{S}_{k,N}+\frac{k^2}{H^2}e^{-2N}\mathcal{S}_k+\frac{M^2}{H^2}
\mathcal{S}_k=0,
\label{eqmot2} \end{eqnarray}
resulting in simple solutions of the following form
\begin{eqnarray}
\mathcal{R}_k(N)&=&e^{-3N/2}\left[C_pJ_{\frac{3}{2}}\left(e^{-N}\frac{k}{H}\right)+C_mJ_{-\frac{3}{2}}\left(e^{-N}\frac{k}{H}\right)\right]
\label{eqmotsol1} \\
\mathcal{S}_k(N)&=&e^{-3N/2}\left[D_pJ_{\frac{1}{2}\sqrt{9-4M^2/H^2}}
\left(e^{-N}\frac{k}{H}\right)+D_mJ_{-\frac{1}{2}\sqrt{9-4M^2/H^2}}
\left(e^{-N}\frac{k}{H}\right)\right].
\label{eqmotsol2} \end{eqnarray}
Initial conditions corresponding to the
Bunch-Davies vacuum are obtained for $C_{p_i}=1$, $C_{m_i}=i$.
For the massive mode, the coefficients must be chosen more carefully, so that they
reproduce the free-wave solution for $N\to - \infty$ when $M$ becomes negligible.
They read
\begin{eqnarray}
 D_{p_i}&=&-\sqrt{2}(1+i)\frac{e^{\frac{i\pi}{4}\sqrt{9-4M^2/H^2}}}{1-e^{i\pi\sqrt{9-4M^2/H^2}}} e^{i\phi}
 \label{coeffm1} \\ D_{m_i}&=&\sqrt{2}(1+i)\frac{e^{\frac{3i\pi}{4}\sqrt{9-4M^2/H^2}}}{1-e^{i\pi\sqrt{9-4M^2/H^2}}} e^{i \phi}.
 \label{coeffm2} \end{eqnarray}
We have also included an arbitrary phase difference $\phi$ between the
curvature and isocurvature modes at early times. This phase may affect the profile 
of the spectra by modifying the interference patterns when
the two modes interact. However, we have found that the quantitative conclusions 
about the spectrum enhancement and the characteristic oscillations it may display
are largely unaffected. For this reason, we use $\phi=0$ in our analysis.

The evolution in the interval $N_1<N<N_2$ is complicated because of the coupling 
between the two modes. The main features are more clearly visible if we neglect the 
expansion of space, which is a good approximation for $\Delta N=N_2-N_1\lta 1$. 
The evolution equations now become
\begin{eqnarray}
\mathcal{R}_{k,NN}+\frac{k^2}{H^2}\mathcal{R}_k+2\etapz\mathcal{S}_{k,N}&=&0
\label{eveqsimple1} \\
\mathcal{S}_{k,NN}+\left(\frac{k^2}{H^2}+\frac{M^2}{H^2}-\etapz^2\right)\mathcal{S}_k
-2\etapz\mathcal{R}_{k,N}&=&0.
\label{eveqsimple2} \end{eqnarray}
Following refs. \cite{Palma:2020ejf, Fumagalli:2020nvq}, we look for solutions of the form 
\be
\mathcal{R}_k=A e^{\omega N}, 
\;\;\;\; 
\mathcal{S}_k=B e^{\omega N}.
\label{ABi} \ee
There are four independent solutions
\be
\omega_i=\pm\frac{1}{\sqrt{2}}\sqrt{-\left(\frac{M^2}{H^2}+3\etapz^2+
2\frac{k^2}{H^2}\right)\pm\sqrt{\left(\frac{M^2}{H^2}+3\etapz^2\right)^2
+16\frac{k^2}{H^2}\etapz^2}},
\label{omegi} \ee
with $i=1,2,3,4$ corresponding to the combinations of signs  $++$, $-+$, $+-$, $--$, 
respectively.
The corresponding values of $B_i$ are
\begin{eqnarray}
B_{1,2}&=&f_+\, \omega_{1,2}\, A_{1,2}
\label{B12} \\
B_{3,4}&=&f_-\, \omega_{3,4}\, A_{3,4},
\label{B34} \end{eqnarray}
with 
\be
f_\pm=\frac{4\,\etapz}{\frac{M^2}{H^2}-5\etapz^2\pm
\sqrt{\left(\frac{M^2}{H^2}+3\etapz^2\right)^2+16\frac{k^2}{H^2}\etapz^2}}.
\label{fpm} \ee
The solutions on either side of $N_1$ and $N_2$ can be matched, assuming the
continuity of $\Rc_k(N)$, $\Sc_k(N)$ and $\Sc_{k,N}(N)$. 
The first derivative of $\Rc_k(N)$ must account for
the $\delta$-function arising from the derivative of $\eta_\perp(N)$ at these 
points. This leads to the conditions
\begin{eqnarray}
\mathcal{R}_{k,N}(N_{1-})&=&\mathcal{R}_{k,N}(N_{1+})+2\etapz\mathcal{S}_k(N_1)
\label{Rder1} \\
\mathcal{R}_{k,N}(N_{2-})&=&\mathcal{R}_{k,N}(N_{2+})-2\etapz\mathcal{S}_k(N_2).
\label{Rder2} \end{eqnarray}

For given initial conditions $C_{p_i},C_{m_i},D_{p_i},D_{m_i}$ before 
the ``pulse", 
one can calculate, through the use of these boundary conditions, 
the constants $A_i$ within the ``pulse", and eventually  
the final coefficients of the free solutions 
$C_{p_f},C_{m_f},D_{p_f},D_{m_f}$ after 
the ``pulse". 
In analogy to the previous section, one 
can thus determine the matrix $M_{\text{pulse}}$ that links the
solutions before and after the ``pulse": 
\be
\begin{bmatrix}C_{p_f}\\C_{m_f}\\D_{p_f}\\D_{m_f}\end{bmatrix}
=M_{\text{pulse}}(N_1,N_2,k,M,\eta_\perp)
\begin{bmatrix}C_{p_i}\\C_{m_i}\\D_{p_i}\\D_{m_i}\end{bmatrix}.
\label{matrix44} \ee
This matrix facilitates calculations for more complex problems with mutiple ``pulses",
occuring when 
the linear valley of the potential is interrupted by multiple, 
successive turns. Unfortunately, the expressions for 
the matrix elements are extensive and not very illuminating. For this reason we do 
not present them explicitly.

The influence of the ``pulse" on the evolution of the perturbations can be inferred
through inspection of eq. (\ref{omegi}). 
Two of the solutions ($\omega_3$ and $\omega_4$) are purely imaginary, resulting in 
oscillatory behaviour. 
The other two ($\omega_1$ and $\omega_2$) 
have a more complicated dependence on the parameters of the problem. 
For $\frac{k}{H}\geq\sqrt{\etapz^2-\frac{M^2}{H^2}}$ 
they are imaginary as well, but for $\frac{k}{H}\leq\sqrt{\etapz^2-\frac{M^2}{H^2}}$
they become real, thus inducing exponential growth or suppression. 
In the limit $\Delta N\rightarrow 0$, $\etapz\to\infty$, with the total area of 
the ``pulse" (or total angle of the turn) $\Delta\theta=\etapz \Delta N$ kept constant, we have 
\be
\omega_{1,2}=\pm \frac{k}{H\sqrt{3}}+\mathcal{O}(\Delta N).
\label{infi}\ee
We expect then that 
the spectrum will be enhanced by a factor 
\be P_\Rc \sim \exp{\left[\frac{2k}{H\sqrt{3}}\Delta N\right]}= \exp{\left[\frac{2\Delta\theta}{\sqrt{3}}\frac{k/H}{\etapz}\right]}.
\label{estenh} \ee
The fact that the maximal turn $\Delta\theta$ cannot exceed $\pi$ for 
canonical kinetic terms implies that 
the enhancement of the spectrum appears for scales 
$k/H$ comparable to $\etapz$. 
As the general solution is a superposition of all independent solutions (\ref{ABi}),
(\ref{omegi}), the exponential growth is accompanied by oscillatory behaviour with
a characteristic frequency set by $\etapz$.
For large $k$ the spectrum returns to its scale-invariant
form, as the effect of the ``pulse" diminishes.

The constraint on $\Delta \theta$ implies that a single turn results only in
limited growth of the spectrum \cite{Palma:2020ejf}. However, multiple turns can 
have an additive, and often resonant, effect. In this respect it is important to
point out another feature of the solutions. 
We are interested in the regime $\etapz\gg M/H$, so that we can neglect the
effect of the mass inside the ``pulse". 
For $k/H\ll \etapz$, strong oscillations, with a frequency set by $\etapz$, occur
within the ``pulse". On the other hand, outside the ``pulse"
the characteristic frequency of oscillations, modulated by the expansion, 
is set by $k$ or $M$. 
The continuity of $\Sc_k(N)$ and its derivative implies that
the amplitude of oscillations increases significantly when the perturbation 
exits the ``pulse". The effect is visible if one matches at $N=0$ 
the toy functions $f_i(N)$ and $f_f(N)$ given by   
\be
f_{i,f}(N)=A_{i,f}e^{-i k_{i,f}N}+B_{i,f}e^{i k_{i,f}N}.
\label{toyt} \ee
This results in  
\begin{eqnarray}
A_f&=&\frac{1}{2}\left(1+ \frac{k_i}{k_f}\right) A_i
+\frac{1}{2}\left(1-\frac{k_i}{k_f}\right) B_i
\nonumber \\
B_f&=&\frac{1}{2}\left(1- \frac{k_i}{k_f}\right) A_i
+\frac{1}{2}\left(1+\frac{k_i}{k_f}\right) B_i.
\label{toymatch} \end{eqnarray}
For $k_i\ll k_f$, as when entering a ``pulse", the constants $A_f$, $B_f$ are
comparable to $A_i$, $B_i$, while for $k_i\gg k_f$, as when exiting a ``pulse",
they are greatly enhanced.
We shall see realizations of this effect in the two-field context through the
numerical solution of eqs. (\ref{RRR}), (\ref{SSS}) in the following.

Before completing the subsection, we point out that 
the alternative formulation of the calculation of the spectrum that 
was introduced in section \ref{alternative} for the 
single field case can be generalized for two fields.
We present the relevant expressions in appendix \ref{appendixb}.

\subsection{Numerical evaluation of the spectra} \label{numerical}

\begin{figure}[t!]
\centering
\includegraphics[width=0.48\textwidth]{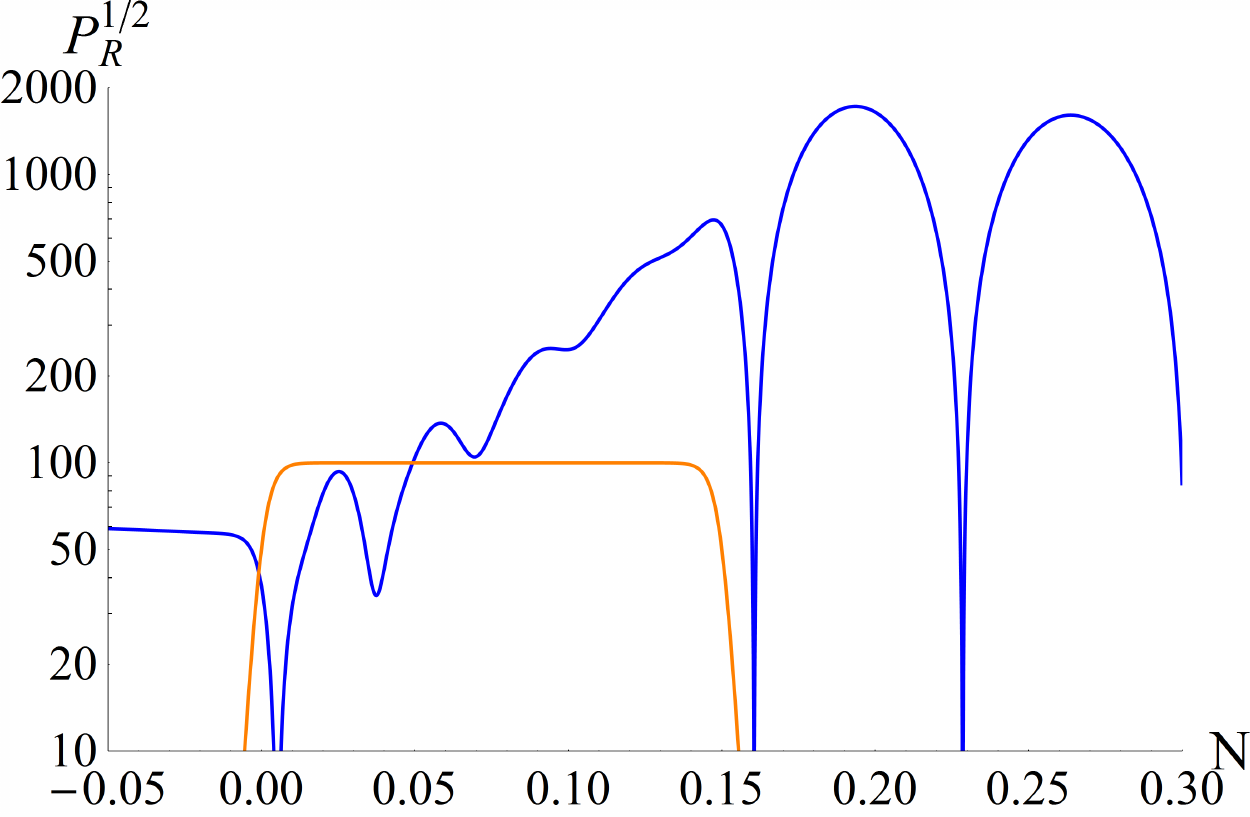} \includegraphics[width=0.48\textwidth]{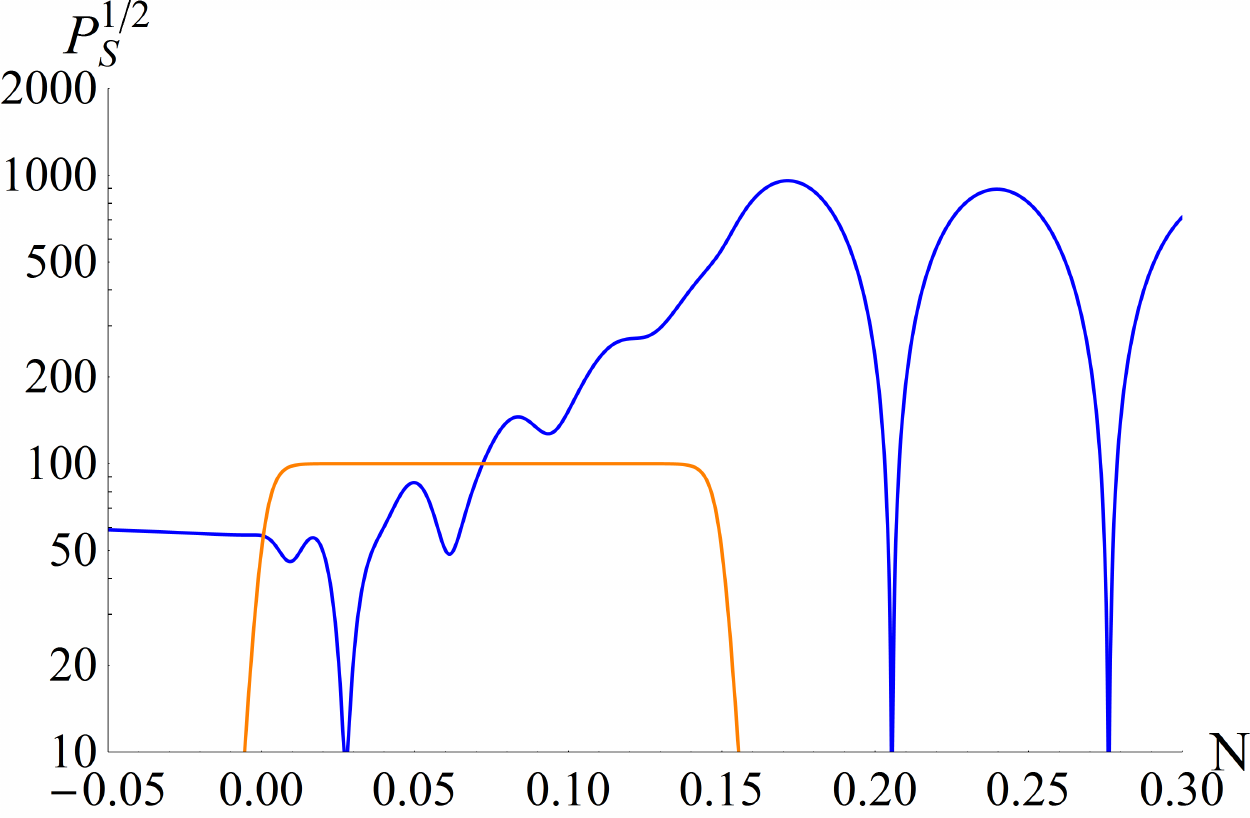}
\\ 
\includegraphics[width=0.48\textwidth]{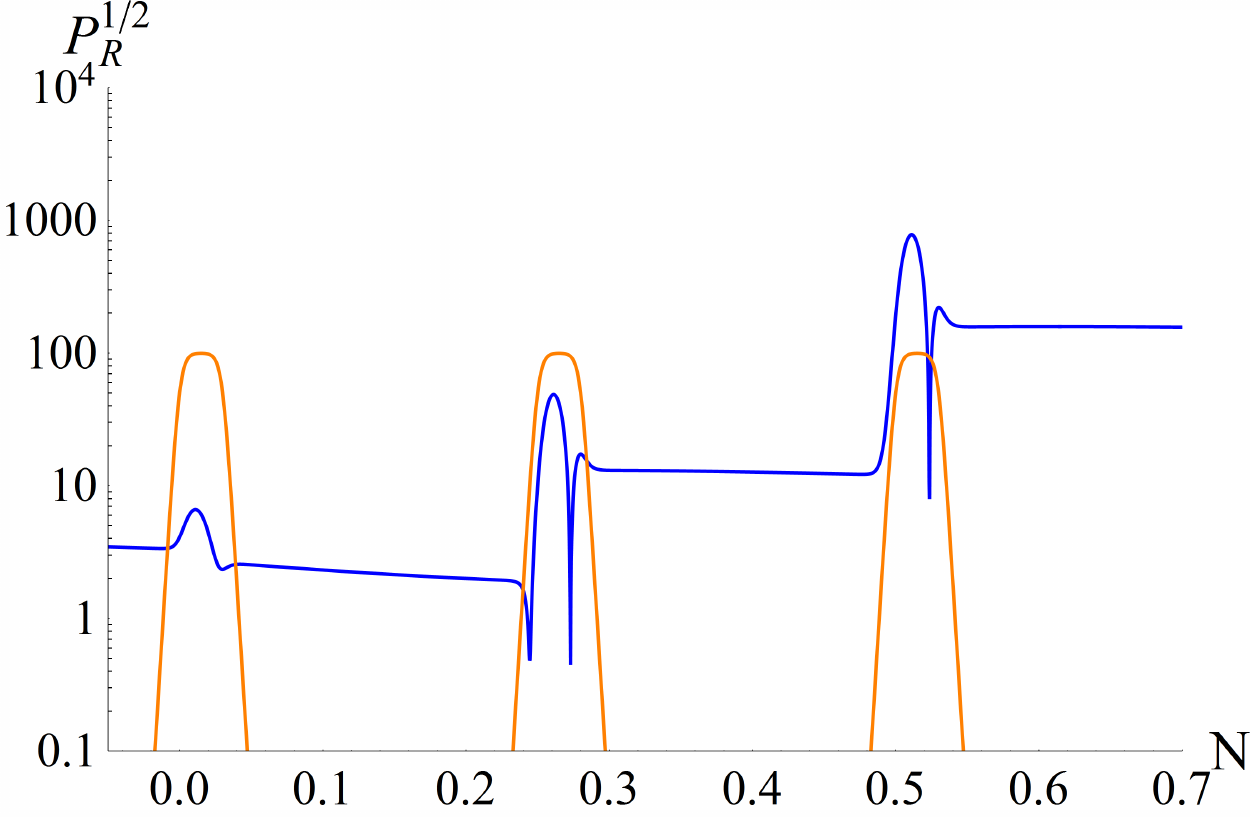} \includegraphics[width=0.48\textwidth]{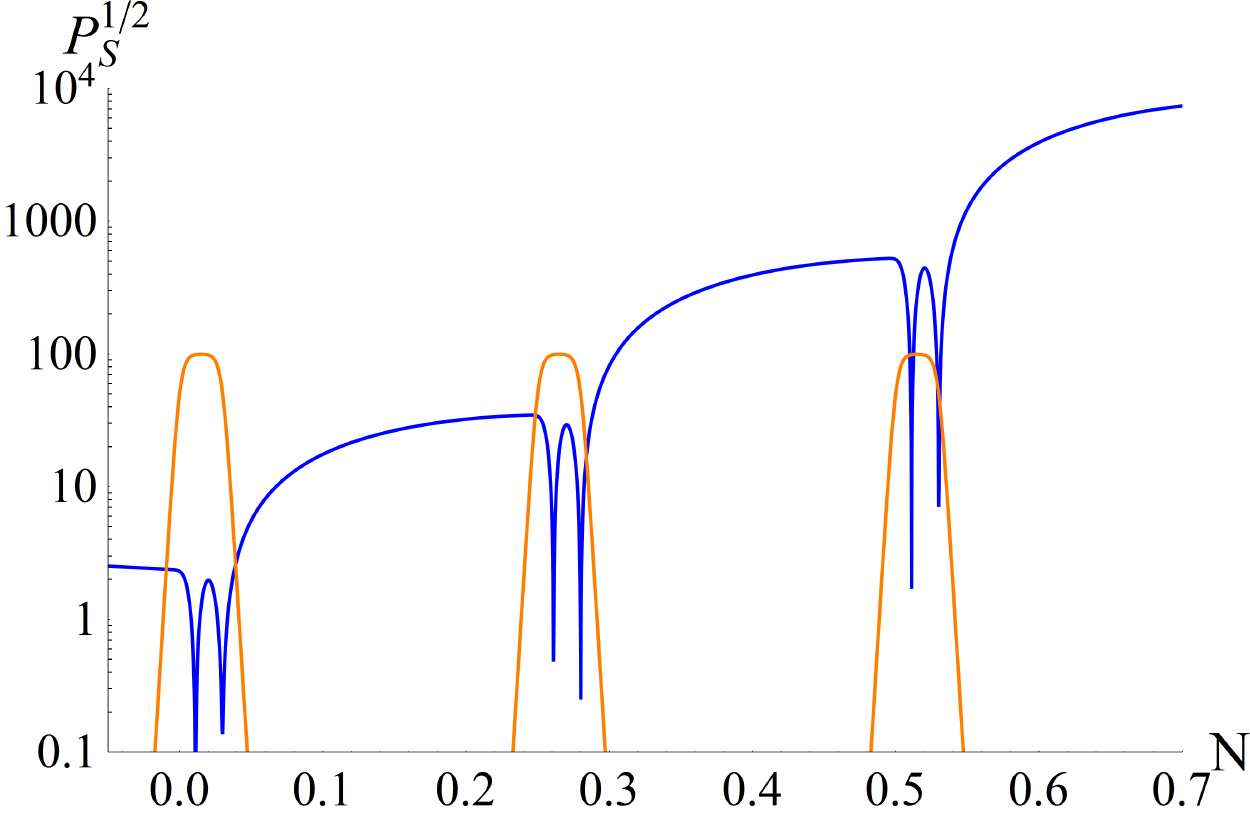}
\caption{
The evolution of the curvature mode (left plots) and isocurvature mode (right plots)
for two types of time dependence for the parameter $\eta_\perp$:
Top row: Thick ``pulse". 
Bottom row: A sequence of three narrow ``pulses" of alternating sign. The
function $|\eta_\perp(N)|$ is depicted. 
The mass of the isocurvature mode is $M/H=5$. The momentum of both modes is
$k/H\simeq 56$ in the top row and $k/H\simeq 3.2$ in the bottom row.
}
\label{figenhance}
\end{figure}

The precise form of the power spectrum in the two-field case is not 
captured easily through an analytic approach, especially when
multiple features appear in the inflaton potential. For this reason, we 
resort to the numerical integration of eqs. (\ref{RRR}), (\ref{SSS}).
In fig. \ref{figenhance} we present the evolution of the 
amplitude of perturbations $\Rc_k(N)$ and $\Sc_k(N)$ in two distinct cases. 
The perturbations are normalized so that the curvature spectrum is equal to
1 for $\eta_\perp=0$.

In the top row of fig. \ref{figenhance}
we present $\Rc_k(N)$ (left plot) and $\Sc_k(N)$
(right plot) for a wide ``pulse", also depicted in the plots. This example does not
represent a physical situation for vanishing curvature ${\mathbb R}$ 
of the internal field manifold, because the total turn $\Delta \theta$
is approximately 15. It is presented in order to demonstrate the
qualitative features discussed in the previous subsection, which become
relevant for ${\mathbb R}\not= 0$. The mode momentum  
is $k/H\simeq 56$, of the same order as the height $\etapz\simeq 100$ of the ``pulse".
The mass of the isocurvature mode is $M/H\simeq 5$. 
One sees clearly the rapid growth of the perturbations inside the ``pulse" and the
appearance of oscillations with frequency set by $\etapz$. After exiting the ``pulse", 
the perturbations oscillate with frequency set by $k$. 
For $N\to \infty$ (a region not depicted in the plots) 
the isocurvature mode asymptotically vanishes because of its 
nonzero mass.
The curvature mode freezes when the horizon is crossed at a value larger than
the one corresponding to the scale-invariant case ($\eta_\perp=0$).

In the bottom row of fig. \ref{figenhance} we present the evolution in the
case of multiple ``pulses". The width of each ``pulse" is much smaller than in the
previous example, so that each turn |$\Delta \theta$| in
field space is smaller than $\pi$. The turns have alternating signs and in the 
plot we depict $|\eta_\perp(N)|$. The mode momentum is $k/H\simeq 3.2$, much 
smaller than $\etapz$. The mass of the isocurvature mode is again $M/H\simeq 5$. 
It is apparent that the growth of both modes within
the ``pulses" is not substantial, even though strong oscillations occur with
frequency set by $\etapz$. The distinctive feature is the strong increase of
the amplitude of the isocurvature mode when the perturbation exits the ``pulse". 
This is expected, according to our discussion at the end of the previous subsection.
The oscillation frequency for this mode outside the ``pulses" is set by the 
mass and is rather low. The location of the ``pulses" is such that a 
resonance effect occurs, with the amplitude of $\Sc_k$ being amplified
each time a ``pulse" is traversed. This effect also triggers the growth of the 
curvature mode. The late time behaviour of both modes (not depicted in the plot)
is similar to the previous case.

\begin{figure}[t!]
\centering
\includegraphics[width=0.48\textwidth]{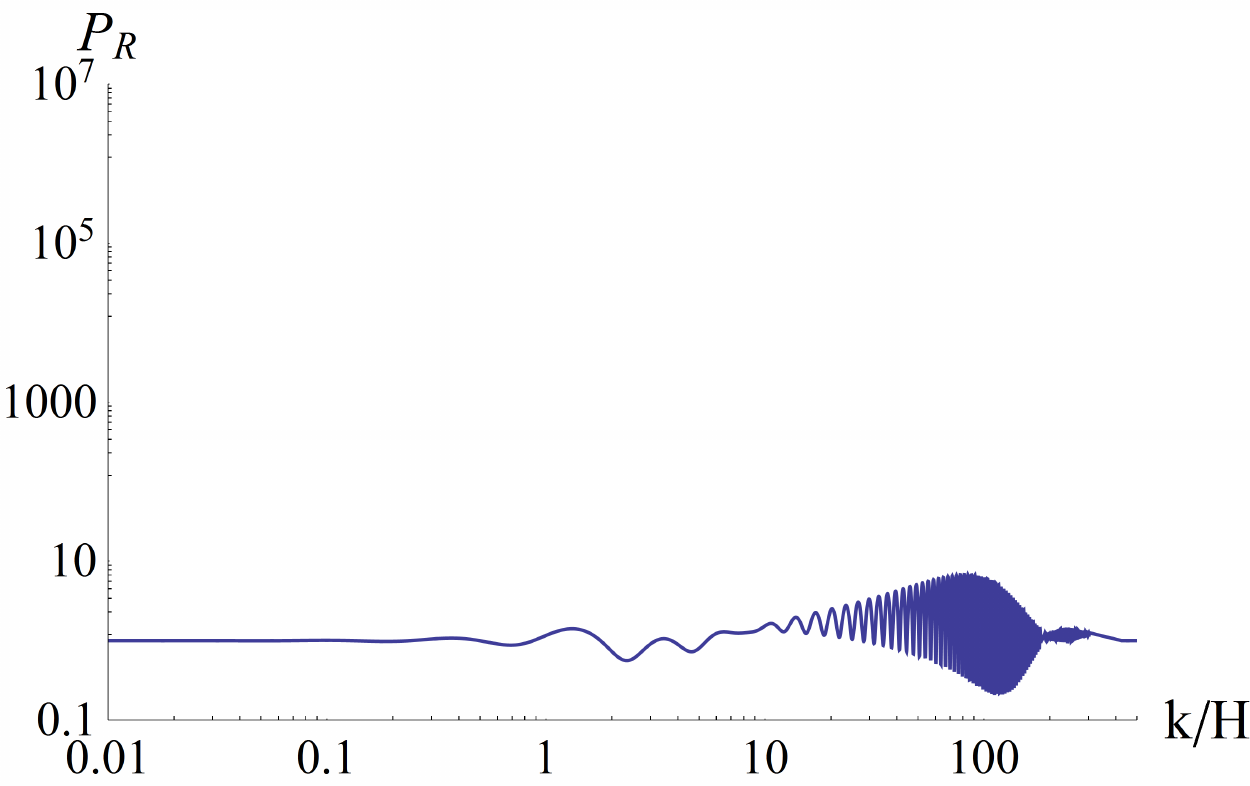} \includegraphics[width=0.48\textwidth]{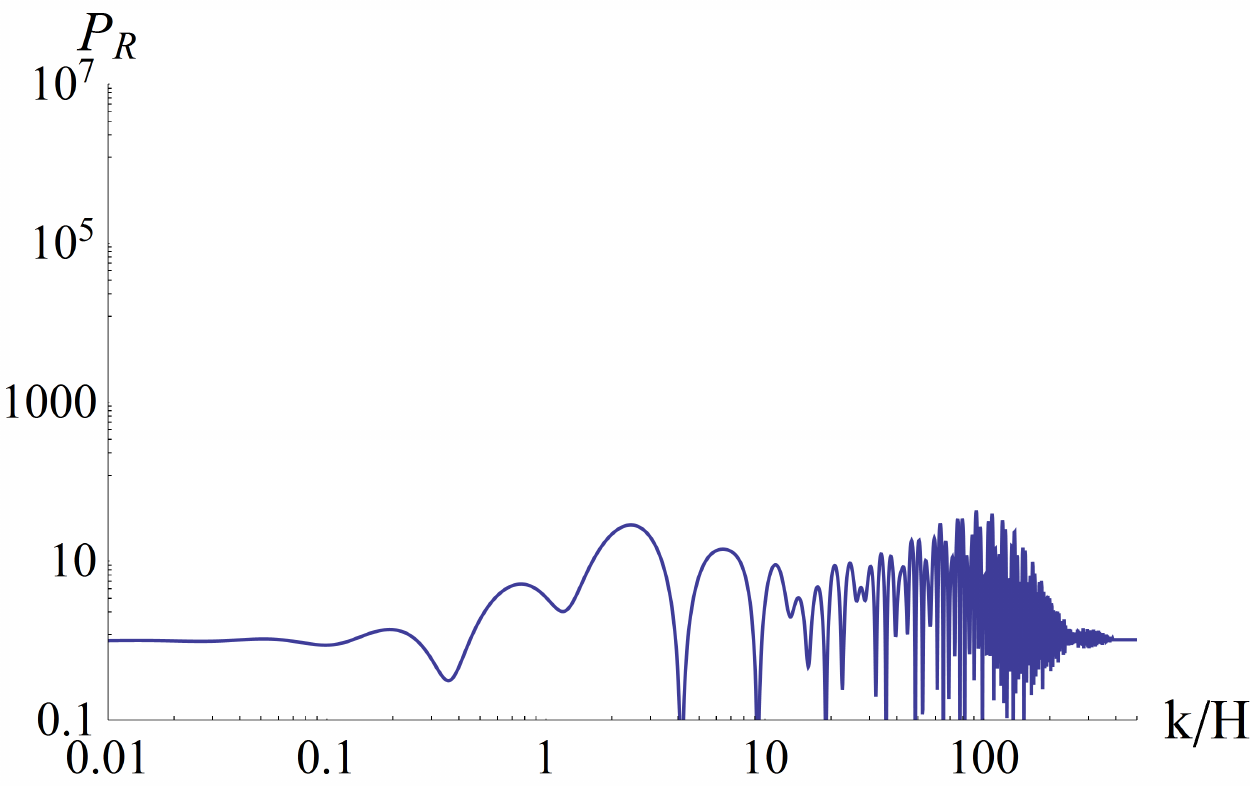} \includegraphics[width=0.48\textwidth]{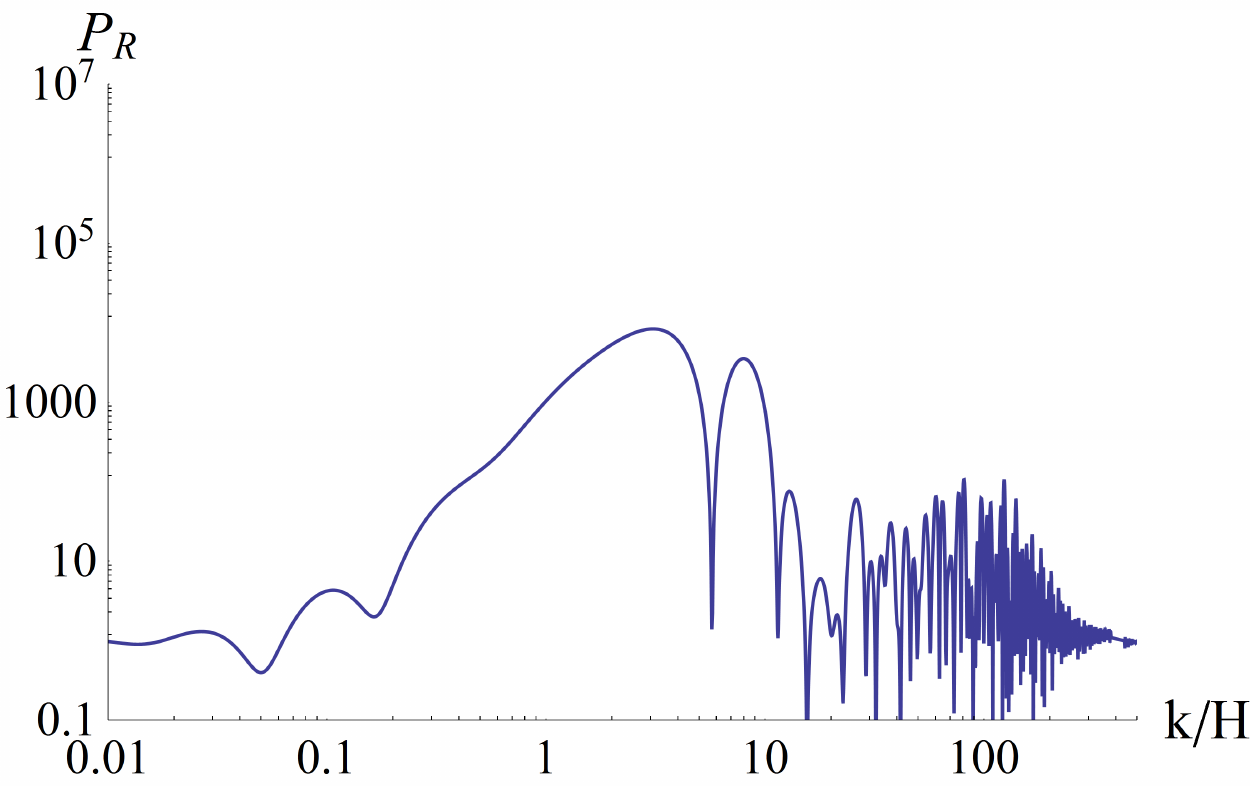} \includegraphics[width=0.48\textwidth]{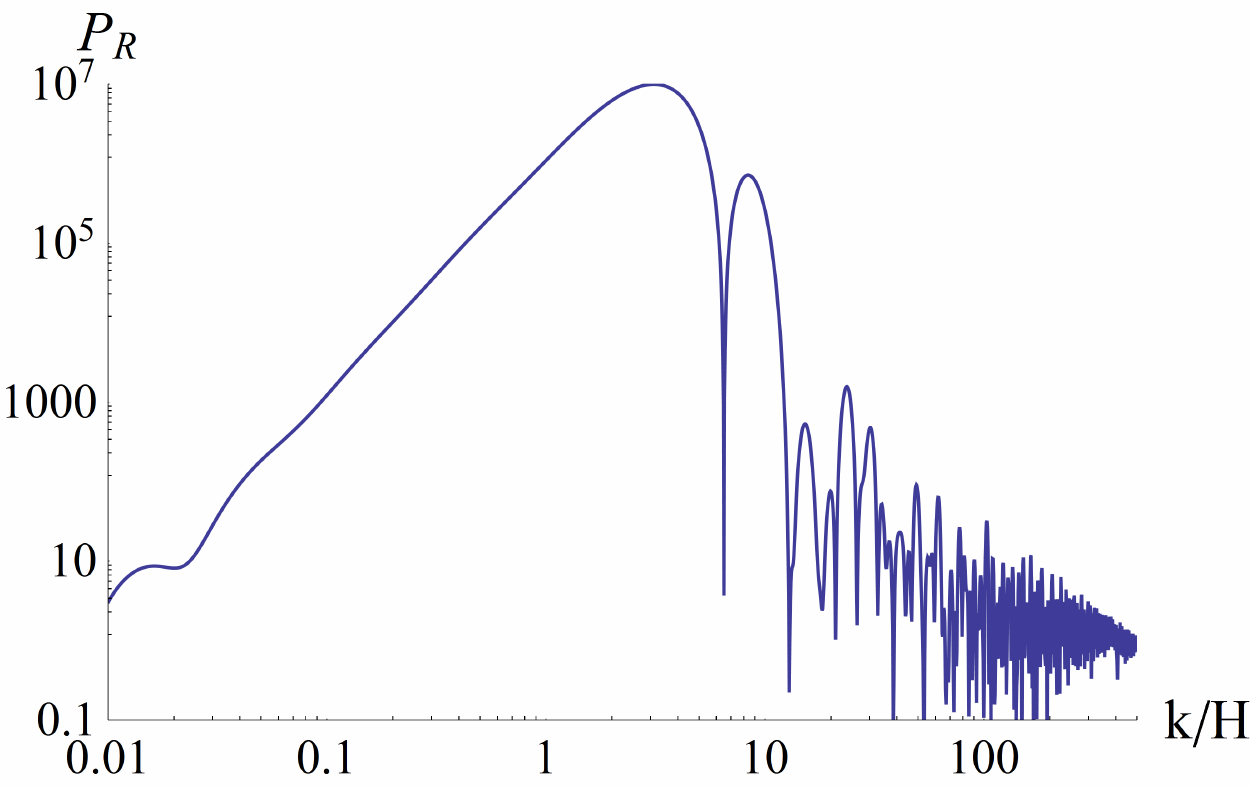}
\caption{
Curvature spectrum induced by a sequence of ``pulses" of alternating signs
in the function $\eta_\perp(N)$. Top row: one ``pulse" (left), two ``pulses" (right).
Bottom row: three ``pulses" (left), four ``pulses" (right). We have kept the same
scales for the axes in all plots for direct comparison.
}
\label{spectrum}
\end{figure}

In fig. \ref{spectrum} we depict the form of the curvature spectrum induced by 
a function $\eta_\perp(N)$ displaying a sequence of ``pulses" such as those in 
the bottom row of fig. \ref{figenhance}. The ``pulses" have alternating signs and are 
located at a distance of 0.25 efoldings from each other. 
Each ``pulse" has a width of approximately
0.03 efoldings, a height $\etapz\simeq 100$ and a total area smaller than $\pi$. 
We have used the same scales in all plots in order to be able to make direct
visual comparisons. The top row includes the spectra generated by 
one (left plot) or two (right plot) ``pulses". 
The bottom row includes the spectra generated by 
three (left plot) or four (right plot) ``pulses".
The spectra are normalized to the scale-invariant one. The location
of the point $k/H=1$ is arbitrary. It corresponds to the mode that crosses
the horizon at the time at which we have set $N=0$ for the number of efoldings,
with $N$ also taking negative values. 
In order to make contact with observations, the normalization of $k$ must be 
set relative to the CMB scale. 

In all plots we observe the enhancement of the spectrum in the region 
$k/H \sim \etapz$. This enhancement increases with the number of ``pulses",
but not to a significant degree. On the other hand,
a strong enhancement appears for multiple ``pulses" at low values of $k/H$,
which increases dramatically with each addition of a ``pulse". This is the
realization of the mechanism that we discussed above 
in relation to the bottom row of fig. \ref{figenhance}. The resonance effect 
appears for $1/k$ comparable to the distance between the ``pulses".  
For our plot we kept this distance fixed. However, similar patterns in the
spectrum are obtained for variable locations of the ``pulses", as long as their
separations are comparable. The exact height of each ``pulse" may also vary.
It must be noted that the effect is strongly amplified by very sharp ``pulses".
We have chosen profiles that match the form of $\eta_\perp(N)$ expected by
models such as the one we discussed in subsection \ref{maximal}.

\section{GWs and the PBH counterpart} \label{GWPBH}

Primordial inhomogeneities imprinted on the CMB are limited to scales of order 
larger than a Mpc. 
Large enhancements of the power spectrum of perturbations on small scales, such as 
those produced through the mechanisms discussed in this paper, 
are not directly visible on the CMB sky  due to photon diffusion damping.
A new method to probe small-scale perturbations is the search for a 
stochastic GW background, see ref. \cite{Domenech:2021ztg} for a recent review. 
The LIGO-Virgo-KAGRA collaboration has already given us some first constraints 
\cite{KAGRA:2021kbb}. 

The stochastic GWs are sourced by the primordial density perturbations. 
Part of the energy stored in the perturbations is released via gravitational radiation
at the moment of horizon reentry.
These are second-order tensor modes generated by first-order scalar modes \cite{Mollerach:2003nq, Ananda:2006af, Baumann:2007zm, Saito:2008jc}.  
The power spectrum of the secondary, or induced, GWs is expressed in a compact 
form as a double integral involving the power spectrum of the curvature 
perturbations:
 \begin{equation}
 \label{tensorPSD2}
     \overline{\mathcal{P}_{h}(\tau,k)}=\int_{0}^{\infty} dt \int_{-1}^{1} ds\ \mathcal{T}(s,t,\tau,k) {\cal P_R} \left(\frac{t+s+1}{2}k\right) {\cal P_R} \left(\frac{t-s+1}{2} k \right)\, .
 \end{equation}
 The overline denotes the oscillation average and ${\cal T}(s,t,\tau,k)$ is a kernel function. In this section we are describing the evolution in terms of 
 the conformal time $\tau$. 
 The variables $t$ and $s$ are 
 defined as  $t=u+v-1$, $s=u-v$, where  $ v={q}/{k}$, $u=|\textbf{k}-\textbf{q}|/{k}$. Further details can be found in ref. \cite{Kohri:2018awv}.
Under radiation domination, most of
 the growth of the induced-GW amplitude occurs rapidly, 
 during a period $\tau_\text{c}\sim O(10)\,k_{\text{peak}}^{-1}$, with 
 $k_{\text{peak}}$ the location of the peak of ${\cal P_R}$.
 After this time 
 the GWs propagate freely.
They have a fractional spectral energy density given by the  GW energy density $\rho_\text{GW}$ relative  to the critical or total energy density $\rho_\text{tot}$ 
per logarithmic wavenumber interval
 \begin{align} \label{OmegaIGWc}
 \Omega_\text{GW}(\tau, k)
 =\frac{1}{\rho_\text{tot}(\tau)} \frac{d \rho_\text{GW}(\tau, k)}{d\ln k} =\frac{1}{24}\left(\frac{k}{a(\tau) H(\tau)} \right)^2 \overline{{\cal P}_h(\tau, k)} \,.
 \end{align} 
At the present time $\tau_0$, the dimensionless spectral density of GWs is
 \begin{align}
 \label{IGW}    
     \Omega_{\textrm{GW}}(\tau_0,f)h^2=0.39\times\left({g_{\ast}\over106.75}\right)^{-{1/3}}\ \Omega_{\text{rad},0} h^2\times\Omega_{\textrm{GW}}(\tau_\text{c},f),
 \end{align}
 where  $\Omega_{\text{rad},0}h^2=4.2\times10^{-5}$, $\tau_0$ the age of the universe today,
  and $\tau_\text{c}$ the time at which most of the GW production takes place.
 The total energy density parameter is obtained by integrating the GW energy density spectrum over the entire frequency interval.
 
 As a rule of thumb, 
 for a scalar spectrum described by a log-normal type distribution
  ${\cal P_R}(k)={\cal P_R} (k_\text{peak})e^{-(z/\sigma)^2}$, 
 with amplitude ${\cal P_R} (k_\text{peak})$ at a narrow peak, 
 $z\equiv\ln(k/k_\text{peak})$ and $\sigma$ the width of the distribution, 
 the  GW spectral energy density today features a peak with amplitude 
 $ \Omega_\text{GW}(\tau_0, f_\text{peak})h^2 \sim 10^{-5}  \sigma^2 {\cal P}^2_{\cal R}(k_\text{peak})$ \cite{Dalianis:2020cla,Pi:2020otn}. 
 A curvature power spectrum with peak amplitude ${\cal P_R}\sim 10^{-2}$ induces a 
 GW peak with $\Omega_\text{GW}(\tau_0, f_\text{peak})\sim 10^{-9} \sigma^{2}$, while 
 ${\cal P_R}\sim 10^{-4}$ induces a GW peak with  
 $\Omega_\text{GW}(\tau_0, f_\text{peak})\sim 10^{-13} \sigma^{2}$. 
 Both values are in the range of sensitivity of designed GW detectors \cite{Thrane:2013oya}, such as LISA \cite{LISA:2017pwj, Barausse:2020rsu} and DECIGO \cite{Kawamura:2020pcg}. The width $\sigma$ of the ${\cal P_R}(k)$ peak 
 is characteristic of each model and shapes the 
 $\Omega_\text{GW}(\tau_0, f)$ spectrum.

 \begin{figure}[!htbp]
   \includegraphics[width=0.51\linewidth]{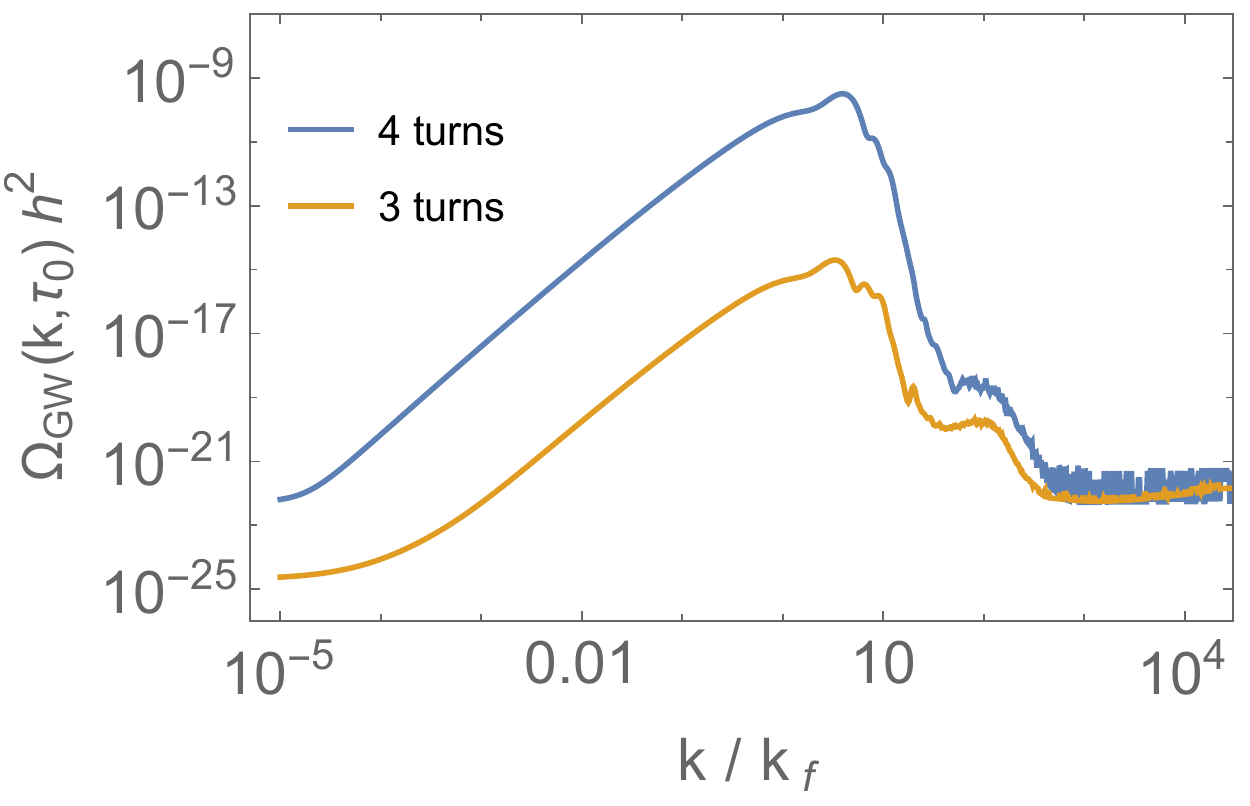}
   \includegraphics[width=0.53\linewidth]{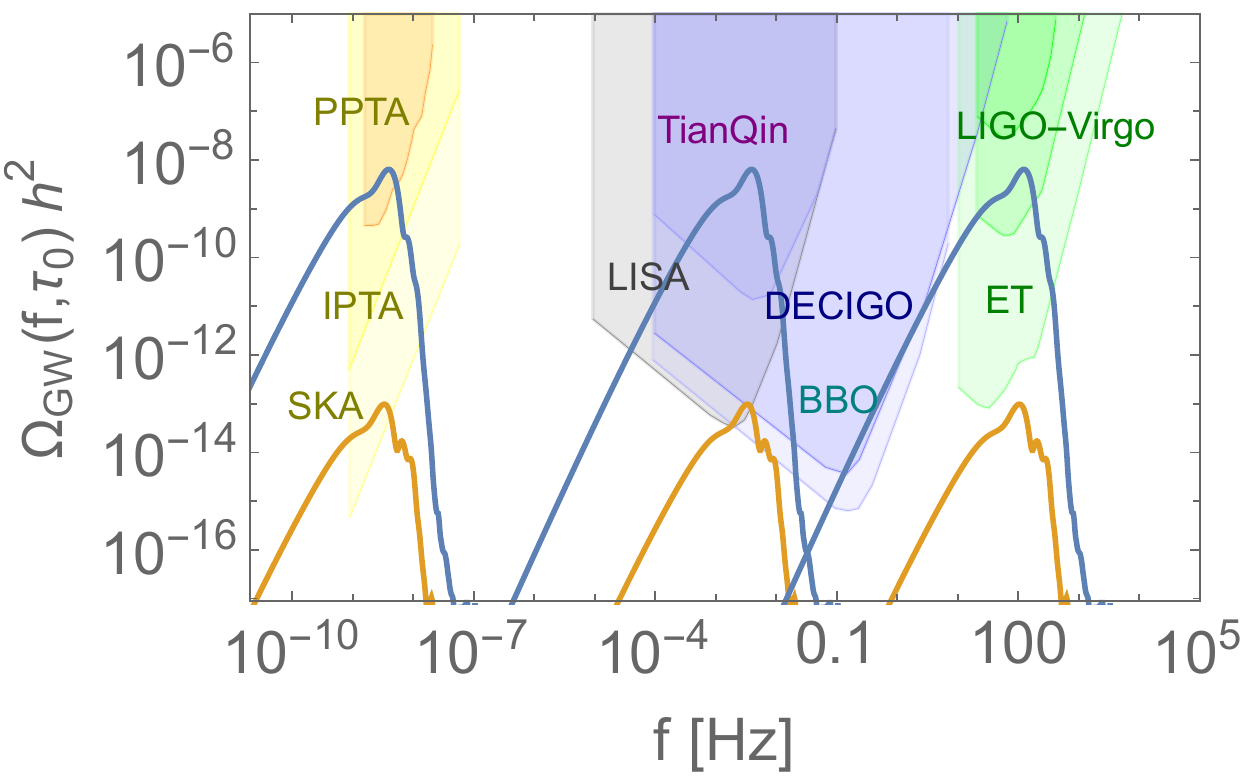}
 \caption{{\it Left plot}: The GW spectral density produced by inflationary 
 trajectories with  3 and 4 successive turns  occurring at the wavenumber $k_f$. 
 The scale $k_f$ is the scale that crosses the horizon at the time when 
 the first feature occurs. In the right plot, 
 we select three different values for
   this scale.
  {\it Right plot}: The GW spectra placed at three benchmark frequencies that lie within the range of sensitivity of current and near future GW experiments.}
  \label{FigGW}
 \end{figure}
 
 \begin{figure}[!htbp]
   \includegraphics[width=0.52\linewidth]{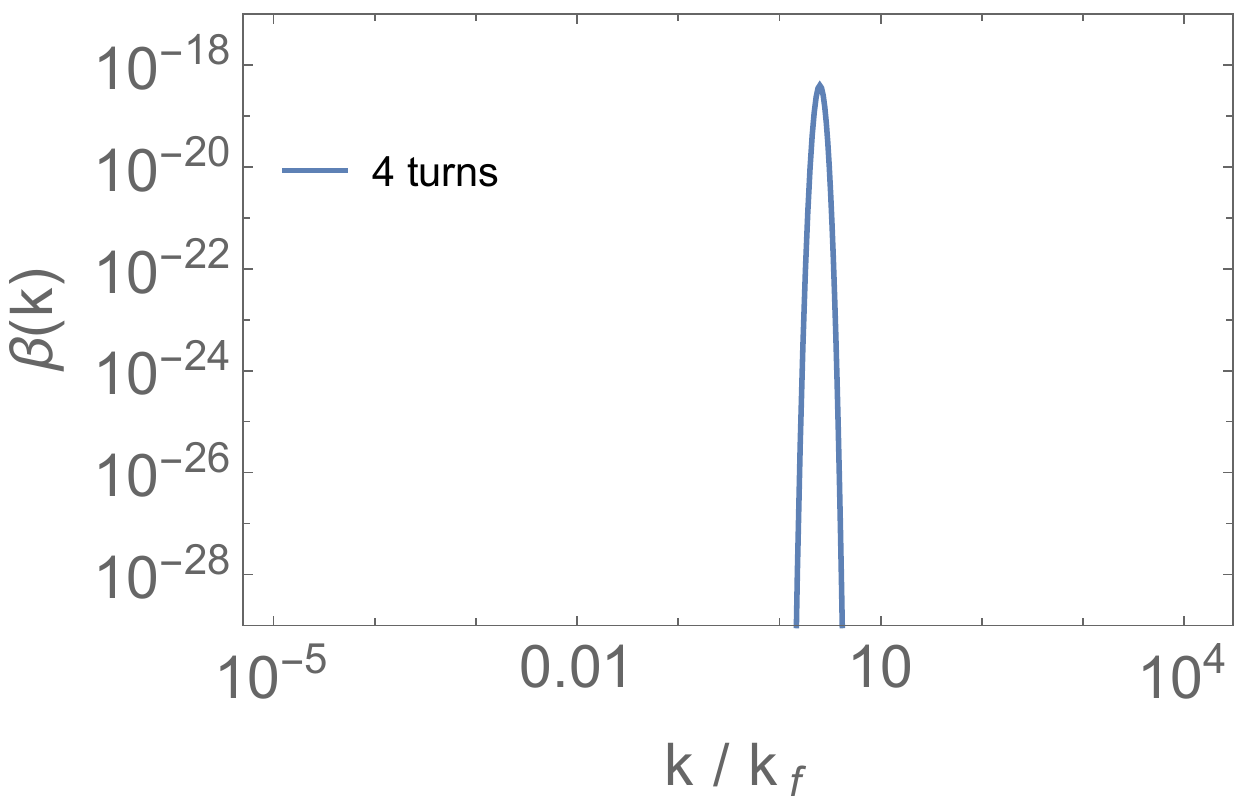}
   \includegraphics[width=0.52\linewidth]{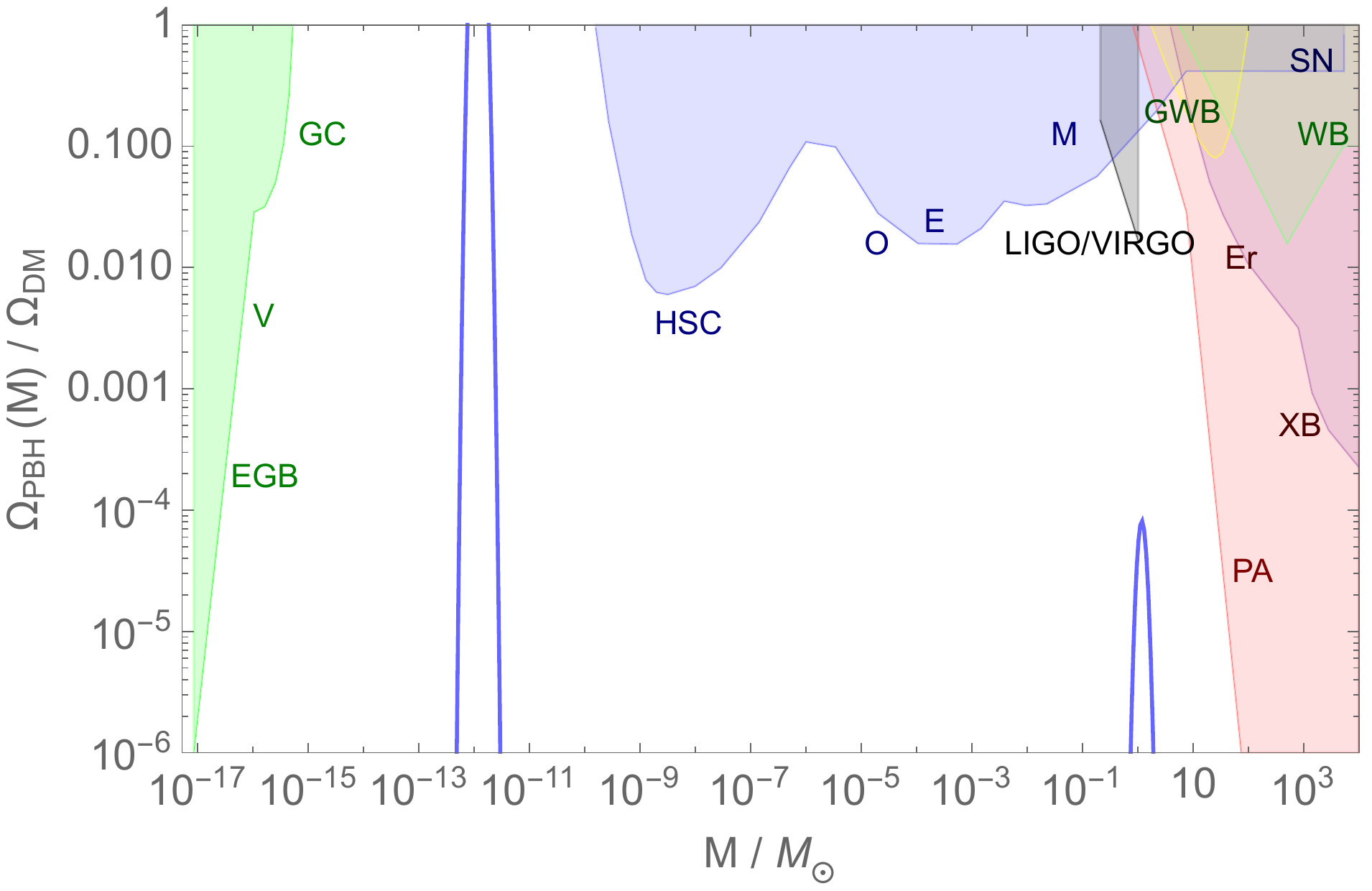}
 \caption{{\it Left plot}: The $\beta$ mass fraction 
  produced by  an inflationary trajectory with four successive turns occurring at the wavenumber $k_f$.  
  {\it Right plot}: The corresponding fractional PBH abundance against the observational constraints \cite{Carr:2020gox}, with masses placed at benchmark scales that can be probed by current and near future GW experiments. These are
  the same scales used for the right plot of fig. \ref{FigGW}.
  The right curve corresponds to the left spectrum in fig. \ref{FigGW}, the left
  curve to the middle spectrum, while the right spectrum in fig. \ref{FigGW}
  corresponds to very small black holes that have evaporated by today.
 }
  \label{FigPBH}
 \end{figure}

 The enhancement of the curvature power spectrum due to several ``pulses" in the 
 evolution of $\eta_{\parallel}$ generates a GW spectrum with a 
 characteristic peak structure, discussed in ref.  \cite{Dalianis:2021iig}. 
 The main peak of the curvature spectrum is found to be narrow, $\sigma<1$, and induces 
 a GW spectrum with a major peak after a flat plateau. Additional oscillatory
 patterns are superimposed on this main peak, reflecting the strong 
 oscillations in the curvature spectrum depicted in fig. \ref{figtwo}. 
 A similar general peak structure is found for a ``pulse" in the evolution  of 
  $\eta_{\perp}$, see Fig. \ref{FigGW}. However, in this case 
  the subleading oscillatory
  patterns  are located after the main peak, which remains fairly smooth. 
This is a consequence of the smooth form of the main peak of the curvature spectrum 
depicted in fig. \ref{spectrum}.

In order to make contact with physical scales we must 
eliminate the ambiguity in the definition of $N$. For this we can use the 
 scale $k_f$ that crosses the horizon at the time $N_f$ when the first feature
 occurs. 
 In the approximation of constant $H$ that we are using, 
 we have $k_f=e^{N_f} H$.  This allows us to eliminate $H$ in favour of $k_f$,
and express every other scale as $k/k_{f} = e^{\delta N}$, 
with $\delta N=N-N_{f}$. In the previous sections we were setting $N_f=0$,
unless stated otherwise. Notice that varying $N_f$ is equivalent to 
varying $k_f$ and allows us to place the peak of the
spectrum at a desired value. As a result, the GW spectrum can be shifted
along the frequency axis, in order to check its detectability depending on
the time of occurrence of the feature that causes the enhancement,
as is done in the right plot of fig. \ref{FigGW}.

 The amplification of the spectrum with respect to the CMB scale depends on the  number 
 of turns, their size and sharpness. 
We considered turns smaller than $\pi$  and 
``pulses"  of amplitude $|\eta_\perp|\lesssim 100$. 
Different configurations can produce an enhancement of 
the curvature spectrum ${\cal P_R}$ of similar size.
For example, a configuration of four turns of angular size 
$\lesssim\pi$ described by four pulses of size  $|\eta_\perp|\sim 100$ 
that last $\Delta N\sim 0.1$  can enhance the power spectrum by a factor of $10^7$. 
A similar enhancement is produced by a configuration of eight turns 
of size $\lesssim \pi/2$ described by eight pulses of size  $|\eta_\perp|\sim 10$.
 Although each configuration produces a different spectral shape, 
 the relative differences are not that evident.  
 The curvature spectrum generally exhibits a major narrow peak, 
 as can be seen in fig.  \ref{spectrum},
 which creates a characteristic  GW spectral peak, common for different configurations.  
The differences appear at $k>k_{\rm peak}$ and correspond to complicated 
oscillatory patterns. These are reflected in the GW spectra of fig. \ref{FigGW} 
in frequency regions of lower amplitude.

Primordial curvature perturbations with large amplitudes at certain scales induce, 
apart from tensor modes, a gravitational collapse of sufficiently 
dense regions that enter the Hubble horizon. 
The mass fraction of PBHs at the formation time, $\beta(M)$, has an acute 
sensitivity to the amplitude of the power spectrum ${\cal P_R}$. 
For the spherically symmetric  gravitational collapse of a fluid with pressure, 
this sensitivity is exponential, 
$\beta \propto \sqrt{\cal P_R}e^{-\delta^2_\text{c}/\cal P_R}$, 
where $\delta_\text{c}$ is the density threshold for PBH formation.  
A sizable PBH abundance is obtained for ${\cal P_R} \sim 10^{-2}$.  
The parameter $\beta$ can increase if the background pressure is decreased, 
as for example during the QCD  phase transition \cite{Jedamzik:1996mr, Byrnes:2018clq}. 
A further decrease of the pressure to vanishing values can affect 
the PBH formation rate significantly \cite{Harada:2013epa, Harada:2016mhb}  
and induce GW signals with different spectral density \cite{Jedamzik:2010hq, Dalianis:2020gup}.

    In our set-up, large amplitudes of ${\cal  P_R}$ can be produced if several 
    strong features, turns or steps, occur during the inflationary trajectory.  
    These features can give an observable result  in the GW channel for  three or more 
    successive turns. In  Fig. \ref{FigGW} we plot the spectral density of the induced 
    GWs for curvature power spectra that have the form depicted in  Fig. \ref{spectrum}. 
     In figure \ref{FigPBH} we plot the corresponding PBH abundance for a threshold value 
     $\delta_\text{c}=0.41$, together with the PBH experimental upper bounds 
     \cite{Carr:2020gox}. 
     For four turns the PBHs produced can constitute a significant fraction of the dark 
     matter in the universe.  
     We recall that during radiation domination the PBH mass spectrum is related to
      the ${\cal P_R}$ peak position $k_\text{peak}$ as $M_\text{PBH}\propto k^{-2}_\text{peak}$, 
      and the PBHs have abundance $\Omega_\text{PBH}\propto \beta M_\text{PBH}^{-1/2}$, see Ref. \cite{Sasaki:2018dmp} for details. 
 PBHs in the mass range $M\sim 10^{-12} M_\odot$ can constitute the entire dark matter in 
 the universe, while the associated GWs will be probed by LISA \cite{elisa} and other future space GW antennas.
 PBHs with mass $M_\text{PBH}\gtrsim M_\odot$ can form binaries and are directly detectable by LIGO-Virgo-KAGRA experiments, while the associated induced GWs can be probed by PTA experiments \cite{Hobbs:2009yy, Janssen:2014dka, Lentati:2015qwp, NANOGrav:2020bcs}. The LIGO-Virgo-KAGRA collaboration already constrains the stochastic GW background  \cite{KAGRA:2021kbb} in the frequency band ${\cal O}(1-10^2)$ Hz, even though PBHs associated with these frequencies are too light to survive in the late universe.

\section{Summary and conclusions} \label{conclusions}

In this paper we extended our previous work \cite{Kefala:2020xsx,Dalianis:2021iig}
on the enhancement of the curvature spectrum during inflation to the two-field case.
The main emphasis of our analysis was on the resonant effect that may occur when
there are several instances with strong 
deviations from approximate scale invariance during the inflationary evolution. 
We did not study a particular model, but looked instead for generic properties 
of the equations of motion for the perturbations which would lead to their 
enhancement. We identified the slow-roll parameter $\eta$ as the quantity
that can trigger the rapid growth of perturbations. In the two-field case
that we analysed in this work, this parameter can be projected onto the 
directions parallel and perpendicular to the trajectory of the 
background fields. The corresponding two components, $\eta_\parallel$ and 
$\eta_\perp$, remain small during most of the evolution, apart from 
short intervals during which they can take large, positive or negative, values.

The typical underlying reason for the appearance of strong features in the
evolution of $\eta_\parallel$ is the presence of points in the inflaton potential
that cannot support slow-roll, such as sharp steps or inflection points.
We focused mostly on the behaviour induced by sharp steps, even though our
formalism can be applied to inflection points as well. 
The acceleration of the background inflaton when a step in the potential
is crossed makes $\eta_\parallel$ attain very negative values initially, followed
by positive values when the inflaton settles in a slow-roll regime on a
plateau following the step. The first part of the evolution is very short for
sharp steps, while the second lasts longer with $\eta_\parallel\simeq 3$. 
During this second part, $\eta_\parallel$ acts as negative friction,
leading to the rapid growth of the curvature perturbation. 

On the other hand, $\eta_\perp$ grows large during sharp turns in
field space. The typical situation involves an inflaton potential that contains 
an almost flat valley, with straight parts interrupted by sharp turns. 
The evolution along the valley satisfies the slow-roll conditions, apart from 
the short intervals during which the fields go through the turns and 
$\eta_\perp$ takes large values, positive or negative, depending on the
direction of the turn. The effect of $\eta_\perp$ is sign-independent 
and twofold: a) it triggers
the strong growth of the perturbation perpendicular to the trajectory (the
isocurvature mode), and b) couples the isocurvature mode to the curvature 
mode along the trajectory, thus inducing its growth as well. 

The focal point of our analysis was the additive effect of several features
leading to the resonant growth of the curvature spectrum. Typical spectra 
arising when the enhancement comes through $\eta_\parallel$ are presented in 
fig. \ref{figtwo}. Three or four strong features are sufficient to
produce an enhancement by six or seven orders of magnitude. A clear
characteristic of the spectrum is the presence of strong oscillations, even around
the peak, which reflect the characteristic times at which the features in the
evolution of $\eta_\parallel$ appear. When the enhancement comes through 
$\eta_\perp$, the resulting spectra have the typical form presented in 
fig. \ref{spectrum}. Again, three or four features in the evolution of
 $\eta_\perp$ are sufficient in order to induce an enhancement of the 
 power spectrum by six or seven orders of magnitude. However, the spectrum
 now displays a smooth main peak, followed by a region of strong oscillatory
 behaviour. This form is a result of the mechanism that leads to the 
 growth of the isocurvature perturbations, discussed in subsections \ref{qualitative}
 and \ref{numerical} and depicted in fig. \ref{figenhance}. It must be pointed out
 that the constructive interference of several features is affected by their 
 separation, as discussed in subsection \ref{numerical}. 
 The resonance effect appears for $1/k$
 comparable to the distance between the “pulses”.

Apart from inducing the CMB anisotropies, inflation is a causal mechanism that 
can seed PBH formation and GW production in the early universe.
Given the fact that we are largely ignorant of the form of the primordial curvature 
power spectrum at small scales, it is legitimate to consider every possibility about 
its amplitude.  Non-minimal scenarios, such 
as those with turns,  generate non-flat shapes with prominent peaks. These can
lead to PBH and GW production, as we discussed in section \ref{GWPBH}.  
In order  that a detectable GW background is produced, e.g. within  the 
sensitivity range of LISA,  three or more successive turns have to occur, while 
a significant PBH abundance requires at least four sharp turns. 
The scenario with several steps in the potential also leads to similar conclusions, 
as has been discussed in detail in refs. \cite{Kefala:2020xsx,Dalianis:2021iig}. 
A comparison of the corresponding GW spectra demonstrates that they both display
a prominent maximum that may be localized
within the range of sensitivity of designed GW detectors 
\cite{Thrane:2013oya}, 
such as LISA \cite{LISA:2017pwj, Barausse:2020rsu} and DECIGO \cite{Kawamura:2020pcg}.
The difference between GW spectra produced through the presence of 
several features in $\eta_\parallel$ and those produced through similar features in 
$\eta_\perp$ lies in the secondary maxima that are present around
the main peak in the first case, while they are absent in the second case. 
Their identification depends crucially on the data resolution of the various 
experiments. However, in principle they provide a means for probing
the mechanism that generates the enhancement of the spectrum.

A big part of our study focused on the attempt to understand the evolution
of the perturbations and the resulting spectra through analytic means.
This is a very difficult task, especially in the presence of several features in
the background evolution, because of the complexity of the problem as it
is reflected in the form of the evolution equations. We employed two 
different approaches:
\begin{itemize} 
\item 
In the first approach we reformulated the problem using
Green's functions, deriving integral equations such as 
eq. (\ref{solg}) for the case of a non-trivial 
$\eta_\parallel$ and eqs. (\ref{solgenn1}), (\ref{solgenn2}) for $\eta_\perp$.
These can be used as a starting point for successive approximations, such
as eqs. (\ref{solg2}) and (\ref{systemsol}), (\ref{CNint}), depicted and discussed
in fig. \ref{figzerozero}. Expressions, such as eq. (\ref{solg2}) provide intuition
on the various frequencies of the oscillations that appear in the power spectrum, 
while eqs. (\ref{systemsol}), (\ref{CNint}) give estimates of the amplitude
of the spectrum far beyond linear perturbation theory. 
Exploiting this approach even further, we reformulated the evolution equations
as a system of differential equations for the coefficients of an expansion of
the general solution in terms of Bessel functions: eq. (\ref{ansatz}) 
in the $\eta_\parallel$ case and
eqs. (\ref{wfw1}), (\ref{wfw2}) in the $\eta_\perp$ case.
This formulation permits the 
analysis of non-minimal initial conditions in a straightforward manner and 
can reduce computational time.
\item
In the second approach, we approximated the features in the evolution of
$\eta$ as square ``pulses" and used appropriate matching conditions at the
beginning and end of each ``pulse" in order to obtain a complete solution.
For the case of a non-trivial $\eta_\parallel$, this approach resulted in 
the approximate expressions derived in subsection
\ref{analyticpulses} and  depicted in fig. \ref{figone}.
For the case of $\eta_\perp$ we used the approximations suggested in refs. 
\cite{Palma:2020ejf, Fumagalli:2020nvq} in order to understand the growth of
the isocurvature perturbations within a single ``pulse" and compare it with 
the situation for a sequence of ``pulses". 

\end{itemize}

The broader picture that emerges from our analysis is that the observable consequences
of inflation can be much more complex than what is suggested by the standard analysis
that assumes small deviations from scale invariance for the whole range of scales 
of the primordial spectrum. The possibility that multiple features may be present in
the background evolution points to a paradigm with richer physical behaviour, 
which is also more natural in the multi-field case. 
It is exciting that the critical examination of such speculations is within 
the reach of experiment.

\section*{Acknowledgments}

The work of I. Dalianis, G. Kodaxis,  and N. Tetradis was supported by the Hellenic Foundation 
for Research and Innovation (H.F.R.I.) under the  “First Call for
H.F.R.I. Research Projects to support Faculty members and Researchers and the procurement of high-cost research equipment grant”  (Project Number: 824).

\appendix

\section{Accuracy of the analytical estimates}
\label{appendixa}

In this appendix we provide an assessment of the 
accuracy of the approximate expressions (\ref{solg2}) and  (\ref{CNint}).
For the matrix $C(N)$, defined in eq. (\ref{CNint}), we
obtain
\begin{eqnarray}
C_{11}(\infty)&=&2\int_{-\infty}^\infty \eta_\parallel(n) \sin\left(e^{-n}\frac{k}{H} \right)
\left[e^{n}\frac{H}{k} \cos\left(e^{-n}\frac{k}{H} \right)+ \sin\left(e^{-n}\frac{k}{H} \right)  
\right] dn
\label{C11} \\
C_{12}(\infty)&=&-2\int_{-\infty}^\infty \eta_\parallel(n)\cos\left(e^{-n}\frac{k}{H} \right)
\left[e^{n}\frac{H}{k}  \cos\left(e^{-n}\frac{k}{H} \right)+ \sin\left(e^{-n}\frac{k}{H} \right)  \right] dn
\label{C12} \\
C_{21}(\infty)&=&2\int_{-\infty}^\infty\eta_\parallel(n) \sin\left(e^{-n}\frac{k}{H} \right)
\left[e^{n}\frac{H}{k} \sin\left(e^{-n}\frac{k}{H} \right)- \cos\left(e^{-n}\frac{k}{H} \right)  
\right] dn
\label{C21} \\
C_{22}(\infty)&=&-2\int_{-\infty}^\infty\eta_\parallel(n) \cos\left(e^{-n}\frac{k}{H} \right)
\left[e^{n}\frac{H}{k}  \sin\left(e^{-n}\frac{k}{H} \right)- \cos\left(e^{-n}\frac{k}{H} \right)  \right] dn,
\label{C22} \end{eqnarray} 
where we have used appropriate partial integrations. 
The first two terms in a Taylor expansion of the exponential give an estimate for $E(\infty)$
that reproduces the approximate result of eq. (\ref{solg2}).
Additional terms account for higher-order corrections.
These improve the convergence for slowly varying functions $\eta_\parallel (N)$.
However, for functions $\eta_\parallel (N)$ that induce an enhancement of the
spectrum by several orders of magnitude, the quantitative accuracy of this
approach is limited. 

In order to check the validity of the approximate expressions  (\ref{solg2})
and (\ref{CNint}),
we consider a sequence of typical patterns of the function $\eta_\parallel(N)$, 
similar to those induced by steep
steps in the inflaton potential \cite{Kefala:2020xsx, Dalianis:2021iig}. 
The form of $\eta_\parallel(N)$ is depicted in 
fig. \ref{figzero} and discussed in the main text.

\begin{figure}[t!]
\centering
\includegraphics[width=0.48\textwidth]{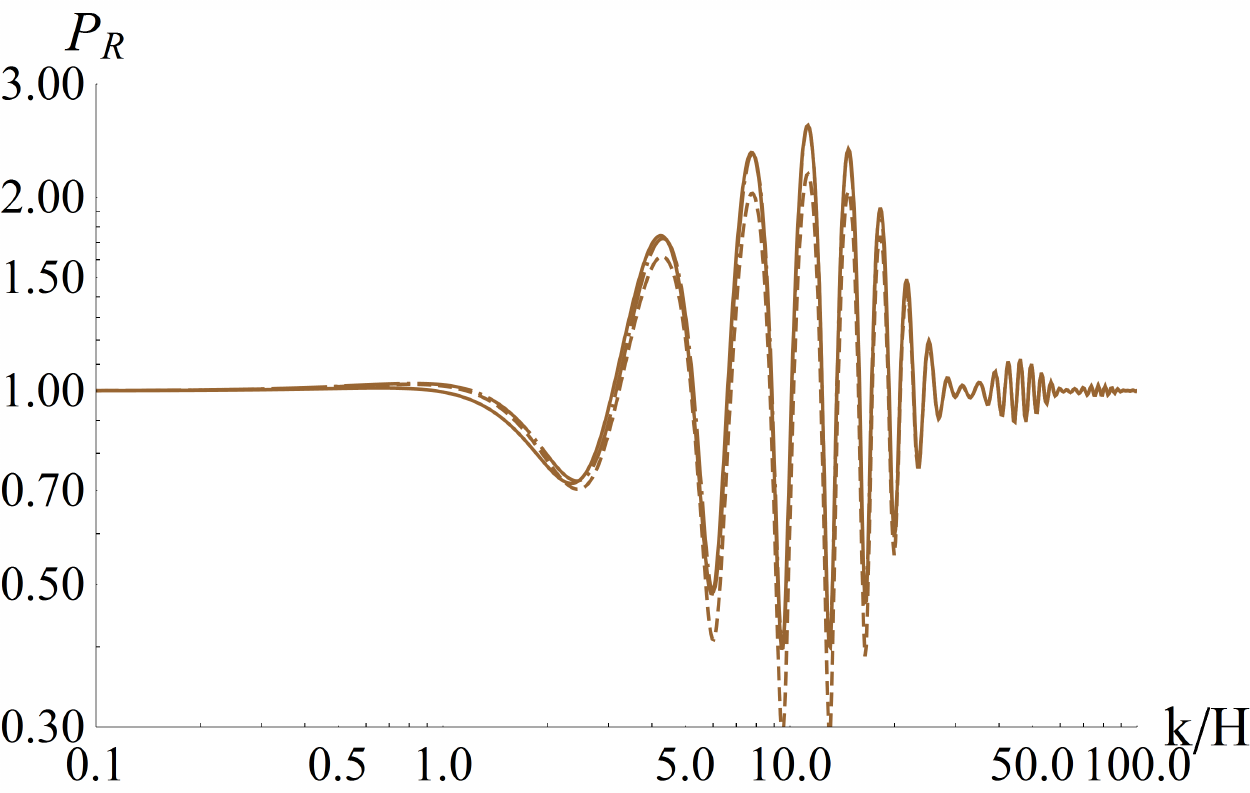} \includegraphics[width=0.48\textwidth]{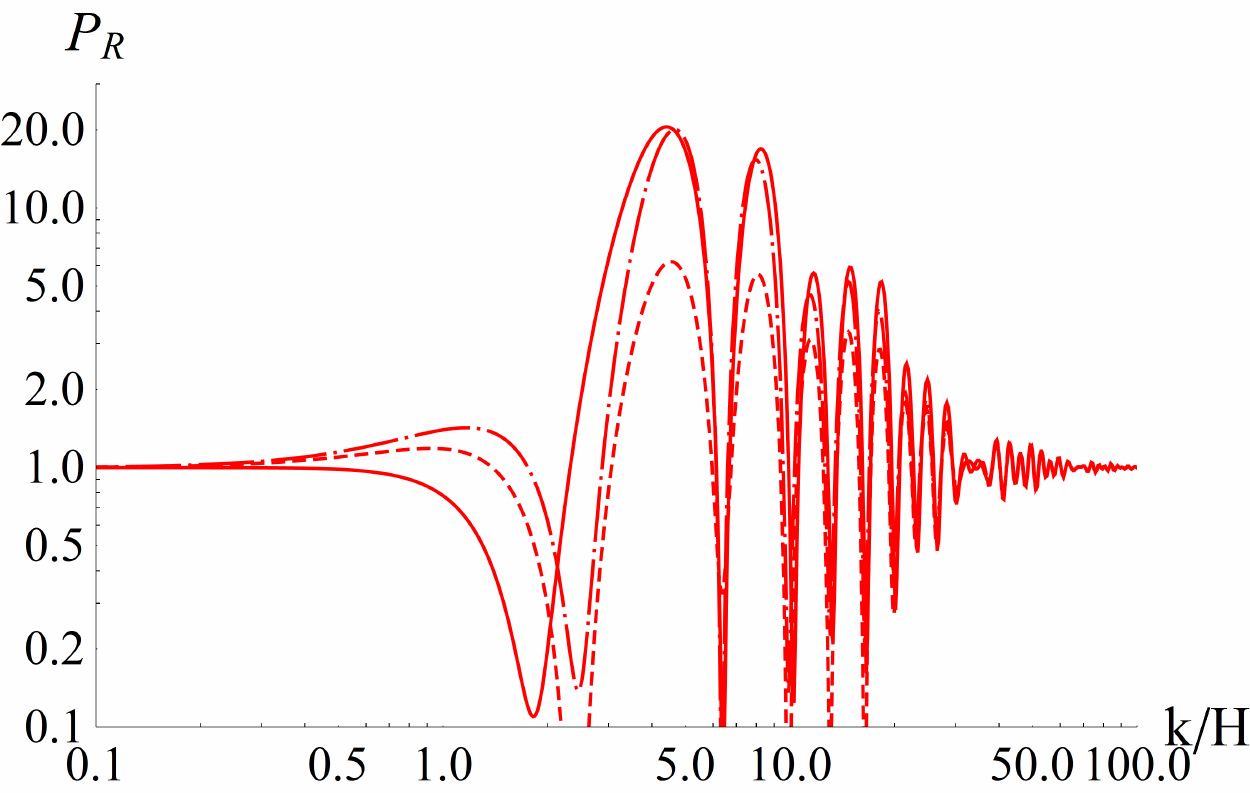} \includegraphics[width=0.48\textwidth]{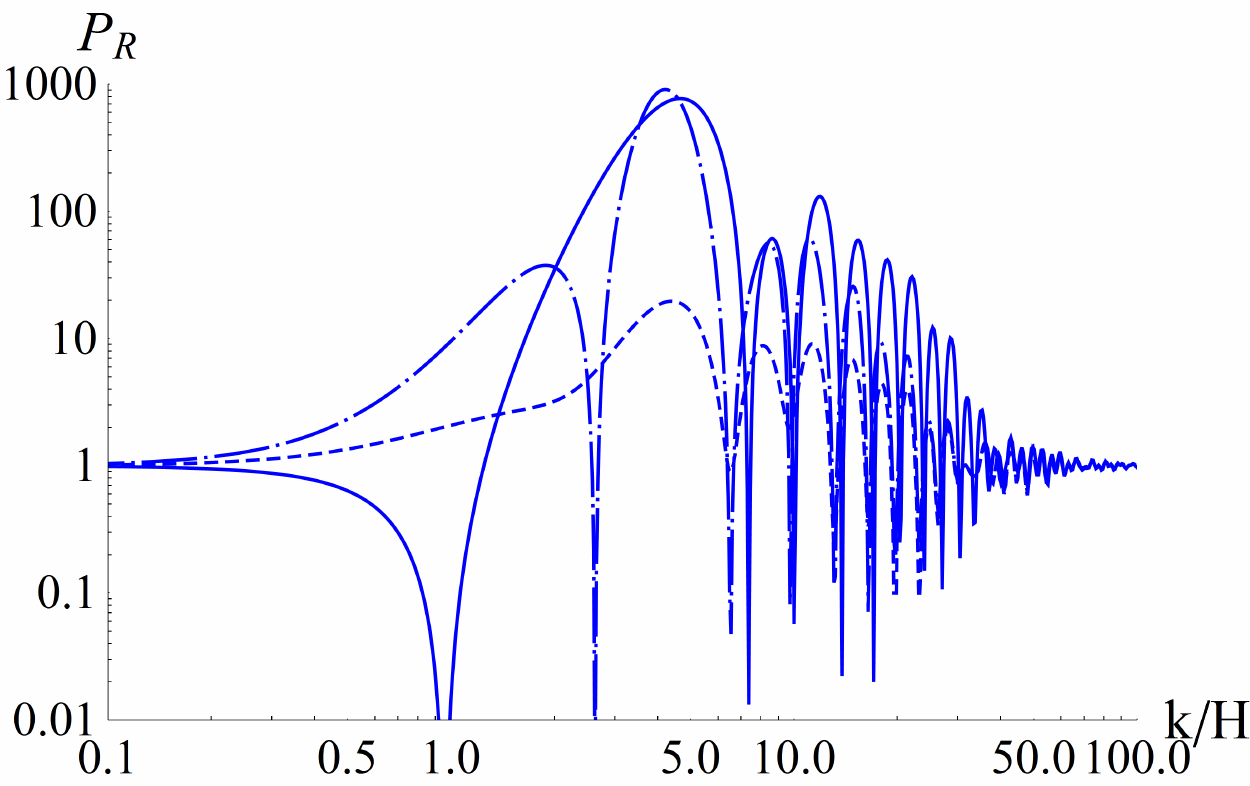} 
\caption{
Power spectra for the three different
smooth curves of fig. \ref{figzero}. The solid curves correspond to the
exact numerical solution of eq. (\ref{RfN}) or the system (\ref{system}). The 
short-dashed curves correspond to the approximation of eq. (\ref{solg2}),
and the long-dashed curves to the approximation of eq. (\ref{CNint}).
Notice the large difference in the vertical scales of the three plots.
}
\label{figzerozero}
\end{figure}

In fig. \ref{figzerozero} we depict the power spectra for the three different
smooth curves of fig. \ref{figzero}. The solid curves correspond to the
exact numerical solution of eq. (\ref{RfN}) or the system (\ref{system}). The 
short-dashed curves correspond to the approximation of eq. (\ref{solg2}),
and the long-dashed curves to the approximation of eq. (\ref{CNint}).
We are interested only in relatively strong deviations from scale invariance, and not
in the absolute normalization of the spectrum. For this reason we have 
assumed that the spectrum is scale invariant at early and late times and 
normalized with respect to its value in these regions. It is apparent from 
fig.  \ref{figzerozero} that both approximations give a very accurate
description of the spectrum when its value is of order 1. When
the spectrum is significantly enhanced both approximations lose accuracy.
However, eq. (\ref{CNint}) gives a reasonable approximation to the 
maximal value of the spectrum and its characteristic frequencies, even for an
enhancement by three orders of magnitude.

\section{An alternative formulation in the two-field case}
\label{appendixb}

The differential equations (\ref{RRR}), (\ref{SSS}) can be turned into integral
equations through the use of the appropriate Green's functions. 
For the curvature mode, the Green's function is given by eq. (\ref{Grless}). 
For the massive isocurvature mode, the generalization is straightforward and 
the retarded Green's function $\bar{G}_k(n,N)$ for $n<N$ is
\be
\bar{G}_{k<}(N,n)=e^{-3N/2}\left(P(n)J_{\frac{1}{2}\sqrt{9-4M^2/H^2}}\left(e^{-N}\frac{k}{H}\right)+Q(n)J_{-\frac{1}{2}\sqrt{9-4M^2/H^2}}\left(e^{-N}\frac{k}{H}\right)\right),
\label{Greeniso} \ee
with 
\begin{eqnarray}
P(n)&=&-\frac{\pi}{2}e^{3n/2}\csc\left(\frac{\pi}{2}\sqrt{9-4M^2/H^2}\right)J_{-\frac{1}{2}\sqrt{9-4M^2/H^2}}\left(e^{-N}\frac{k}{H}\right)
\label{greeniso1} \\
Q(n)&=&\frac{\pi}{2}e^{3n/2}\csc\left(\frac{\pi}{2}\sqrt{9-4M^2/H^2}\right)J_{\frac{1}{2}\sqrt{9-4M^2/h^2}}\left(e^{-N}\frac{k}{H}\right).
\label{greeniso2} \end{eqnarray}
Hence, the general solution can be expressed as:
\begin{eqnarray}
\mathcal{R}_k(N)&=&\bar{\mathcal{R}}_k(N)-2\int_{-\infty}^N G_{k<}(N,n)\left(\frac{\partial}{\partial n}(\eta_\perp(n)\mathcal{S}_k(n))+3\eta_\perp(n)\mathcal{S}_k(n)\right)dn
\label{solgenn1} \\
\mathcal{S}_{k}(N)&=&\bar{\mathcal{S}}_{k}(N)+\int_{-\infty}^N\bar{G}_{k<}(N,n)\left(\eta^2_\perp(n)\mathcal{S}_{k}(n)+2\eta_\perp(n)\mathcal{R}_{k,n}(n)\right),
\label{solgenn2} \end{eqnarray}
\newline
where $\bar{\mathcal{S}}_{k}(N)$, $\bar{\mathcal{R}}_{k}(N)$ are the homogeneous solutions.

The form of the above equations suggests the ansatz
\begin{eqnarray}
\mathcal{R}_k(N)&=&e^{-3N/2}\left(D(N)J_{3/2}\left(e^{-N}\frac{k}{H}\right)+E(N)J_{-3/2}\left(e^{-N}\frac{k}{H}\right)\right)
\label{wfw1} \\
\mathcal{S}_k(N)&=&e^{-3N/2}\left(K(N)J_{\frac{1}{2}\sqrt{9-4M^2/H^2}}\left(e^{-N}\frac{k}{H}\right)+L(N)J_{-\frac{1}{2}\sqrt{9-4M^2/H^2}}\left(e^{-N}\frac{k}{H}\right)\right)
\label{wfw2} \end{eqnarray}
By substituting in the derivative of eqs. (\ref{solgenn1}), (\ref{solgenn2}),
and matching the coefficients of 
the Bessel functions, we obtain a system of four first-order differential equations:
\be
\frac{\partial}{\partial N}\begin{bmatrix}D(N)\\E(N)\\K(N)\\L(N)\end{bmatrix}=F(N)\begin{bmatrix}D(N)\\E(N)\\K(N)\\L(N)\end{bmatrix}
\label{inteq4d}\ee
where $F(N)$ is a $4\times 4$ matrix with nonzero elements $F_{ij}(N)$ given by
\begin{eqnarray}
F_{11}(N)&=&F_{12}(N)=F_{21}(N)=F_{22}(N)=0
\nonumber \\
F_{13}(N)&=&2e^{-3N/2}\eta_\perp(N)\left(A'(N)-3A(N)\right)J_{\frac{1}{2}\sqrt{9-4M^2/H^2}}\left(e^{-N}\frac{k}{H}\right)
\nonumber \\
F_{14}(N)&=&2e^{-3N/2}\eta_\perp(N)\left(A'(N)-3A(N)\right)J_{-\frac{1}{2}\sqrt{9-4M^2/H^2}}\left(e^{-N}\frac{k}{H}\right)
\nonumber \\
F_{23}(N)&=&2e^{-3N/2}\eta_\perp(N)\left(B'(N)-3B(N)\right)J_{\frac{1}{2}\sqrt{9-4M^2/H^2}}\left(e^{-N}\frac{k}{H}\right)
\nonumber \\
F_{24}(N)&=&2e^{-3N/2}\eta_\perp(N)\left(B'(N)-3B(N)\right)J_{-\frac{1}{2}\sqrt{9-4M^2/H^2}}\left(e^{-N}\frac{k}{H}\right)
\nonumber \\
F_{31}(N)&=&-2e^{-3N/2}\frac{d}{dN}\left(\eta_\perp(N)P(N)\right)J_{\frac{3}{2}}\left(e^{-N}\frac{k}{H}\right)
\nonumber \\
F_{32}(N)&=&-2e^{-3N/2}\frac{d}{dN}\left(\eta_\perp(N)P(N)\right)J_{-\frac{3}{2}}\left(e^{-N}\frac{k}{H}\right)
\nonumber \\
F_{33}(N)&=&e^{-3N/2}\eta^2_\perp(N)P(N)J_{\frac{1}{2}\sqrt{9-4M^2/H^2}}\left(e^{-N}\frac{k}{H}\right)
\nonumber \\
F_{34}(N)&=&e^{-3N/2}\eta^2_\perp(N)P(N)J_{-\frac{1}{2}\sqrt{9-4M^2/H^2}}\left(e^{-N}\frac{k}{H}\right)
\nonumber \\
F_{41}(N)&=&-2e^{-3N/2}\frac{d}{dN}\left(\eta_\perp(N)Q(N)\right)J_{\frac{3}{2}}\left(e^{-N}\frac{k}{H}\right)
\nonumber \\
F_{42}(N)&=&-2e^{-3N/2}\frac{d}{dN}\left(\eta_\perp(N)Q(N)\right)J_{-\frac{3}{2}}\left(e^{-N}\frac{k}{H}\right)
\nonumber \\
F_{43}(N)&=&e^{-3N/2}\eta^2_\perp(N)Q(N)J_{\frac{1}{2}\sqrt{9-4M^2/H^2}}\left(e^{-N}\frac{k}{H}\right)
\nonumber \\
F_{44}(N)&=&e^{-3N/2}\eta^2_\perp(N)Q(N)J_{-\frac{1}{2}\sqrt{9-4M^2/H^2}}\left(e^{-N}\frac{k}{H}\right).
\label{44system} \end{eqnarray}
This system of equations must be solved with initial conditions at
$N\to -\infty$. For the Bunch-Davies vacuum $(D,E)=(1,i)$, while 
$(K,L)$ are given by eqs. (\ref{coeffm1}), (\ref{coeffm2}).


\end{document}
